\documentclass[prd,aps,twocolumn,nofootinbib,10pt]{revtex4}
\usepackage{amsmath}
\usepackage{color,graphicx,slashed} 

\begin{document}

\title{Light-cone distribution amplitudes of  a light baryon in large-momentum effective theory}

\author{Zhi-Fu Deng}
\author{Chao Han}
\author{Wei Wang} 
\author{Jun Zeng~\footnote{Corresponding author: zengj@sjtu.edu.cn}} 
\author{Jia-Lu Zhang} 
\affiliation{ 
$^1$ INPAC, Key Laboratory for Particle Astrophysics and Cosmology (MOE),  Shanghai Key Laboratory for Particle Physics and Cosmology, School of Physics and Astronomy, Shanghai Jiao Tong University, Shanghai 200240, China
}

\author{ }
\affiliation{}

\begin{abstract}
Momentum distributions of quarks/gluons inside a light baryon in a hard exclusive process are encoded in the light-cone distribution amplitudes (LCDAs). In this work, we point out that the leading twist  LCDAs of a light baryon can be obtained through a simulation of a quasi-distribution amplitude calculable on lattice QCD within the framework of the large-momentum effective theory.  We calculate the one-loop perturbative contributions to LCDA and quasi-distribution amplitudes and explicitly demonstrate the factorization of quasi-distribution amplitudes at the one-loop level. Based on the perturbative results, we derive the matching kernel in the $\overline{\rm MS}$ scheme and regularization-invariant momentum-subtraction scheme. Our result provides a first step to obtaining the LCDA from first principle lattice QCD calculations in the future.
\end{abstract}

\maketitle

\section{Introduction}

Light-cone distribution amplitudes (LCDAs) of a light baryon describe the momentum distributions of a quark/gluon in a baryonic system and are a  fundamental non-perturbative input in QCD factorization for an exclusive process with a large momentum transfer. An explicit example of this type is weak decays of bottom baryons which are valuable to extract the CKM matrix element in the standard model~\cite{LHCb:2015eia} and to probe new physics beyond the standard model~\cite{LHCb:2017slr}. In addition, in contrast with parton distribution functions that encode the probability density of parton momenta in hadrons, the LCDAs offer a probability amplitude description of the partonic structure of hadrons, from which one can potentially calculate various quark/gluon distributions. Thus the knowledge of LCDAs is also key to understanding the internal structure of light baryons, such as a proton.

Though many progresses have been made in obtaining the LCDAs of a nucleon in the past decades~\cite{Chernyak:1984bm,King:1986wi,Chernyak:1987nu,Chernyak:1987nv,Braun:2014wpa,RQCD:2019hps,Stefanis:1992nw,Bolz:1996sw,Groote:1997yr,Braun:2000kw,Braun:2006hz,QCDSF:2008qtn,Anikin:2013aka,Kim:2021zbz},  most of the available analyses are limited to the few lowest moments of LCDAs. Due to the lack of a complete knowledge of baryon LCDAs, many phenomenological analyses adopt model paramterizations resulting in uncontrollable errors in theoretical predictions for  decay branching fractions of heavy baryons \cite{Lu:2009cm,Huang:2022lfr,Han:2022srw}.   Thus, it is highly indispensable to develop a method to calculate the full shape of baryon LCDAs from the first principle of QCD.

Since LCDAs are defined as the correlation functions of lightcone operators inside a hadron, these quantities can not be directly evaluated on the lattice. In 2013, a very inspiring approach was proposed to circumvent this problem and is now formulated as the large-momentum effective theory (LaMET)~\cite{Ji:2013dva,Ji:2014gla}.  In LaMET, instead of directly calculating light-cone correlations, one can start from equal-time correlations in a large-momentum hadron state, which are known as quasi-distributions. The quasi-distributions share the same infrared properties with lightcone distributions and are connected to PDFs and LCDAs via a matching scheme. Under the framework of LaMET,  encouraging results are recently obtained on the lattice and for recent reviews please see Refs.~\cite{Cichy:2018mum,Zhao:2018fyu,Ji:2020ect} and many references therein.  Based on this approach, results on LCDAs of light mesons can be found in  Refs.~\cite{Zhang:2017bzy,Zhang:2017zfe,Zhang:2020gaj,Hua:2020gnw,LatticeParton:2022zqc,Gao:2022vyh}.  Other methods to extract lightcone PDFs and LCDAs can also be found in Refs.~\cite{Orginos:2017kos,Radyushkin:2017cyf,Ma:2014jla,Ma:2017pxb}.

In this work, we aim to provide an exploration of  the leading twist lightcone distribution amplitude of a light baryon in LaMET. Taking the $\Lambda$ baryon as an example,  we first calculate the one-loop perturbative QCD contributions to LCDAs and quasi-DAs of a light baryon. We demonstrate that these two quantities have the same infrared structure which explicitly validates the factorization at the one-loop level. We also provide an analysis based on expansion by region, which gives direct proof. Based on the one-loop results, we derive the matching kernel. To regularize remnant UV divergences, we also give the matching results in a regularization-invariant momentum-subtraction scheme.  Future improvements in lattice realization will be briefly mentioned in the end.

The rest of this paper is organized as follows. In Sec.~\ref{sec:LCDA}, we present a brief review of the twist-2 LCDAs of a light baryon and the one-loop perturbative results. In Sec.~\ref{sec:Quasi-DA}, we calculate the contributions to the quasi-DA in the modified minimal subtraction scheme. In Sec.~\ref{sec:off_shell}, we calculate the one-loop contributions to quasi-DA with the off-shell external states with a RI/MOM subtraction. In Sec.~\ref{sec:matching}, we give the one-loop matching coefficients from quasi-DA to LCDA. A summary is presented in Sec.~\ref{sec:summary}. Some details are provided in the appendix. 

\section{LCDA at one loop level}
\label{sec:LCDA}

In the factorization analysis of heavy-to-light baryonic  transition, one is led at leading-twist to the matrix element of a three-quark operator between
the vacuum and the baryon state. Taking  the $\Lambda$ baryon which is made of $uds$ as an example,  one can see that the   LCDA is defined by the non-local light-ray operators
 \begin{eqnarray}
 \label{eq:definitionLCDA}
 \epsilon_{ijk}  \langle 0|u_{i}^T(t_1 n) \Gamma d_{j}(t_2 n)  s_{k}(0) |\Lambda\rangle,
 \end{eqnarray}
 with $i,j,k$ being color indices. $T$ denotes the transpose in the spinor space. Under the assignment of  $\bar n$ as the light quark flight direction, the three light quarks are separated in the $n$ direction in coordinate space. The two lightcone unit vectors are defined as $n^\mu=(1,0,0,-1)/\sqrt{2}$ and $\bar{n}^\mu = (1,0,0,1)/\sqrt2$. The covariant derivative is $D_\mu=\partial_{\mu}-igA_{\mu}$.

 Two pieces of gauge links are not shown in the above formulae
\begin{eqnarray}\label{eq:wilsonline}
  \mathcal{W}_{ij}(0,x)= \mathcal{P}{\rm exp}\left[ig_s \int_{x}^0 dt n_\mu A^{\mu}_{ij}(tn)\right].
\end{eqnarray}
It is worthwhile pointing out that the above form of the Wilson line is not unique, but a gauge invariant building block, e.g. for a quark field with color $i$, is 
\begin{eqnarray}
Q_{i}(x)= \mathcal{W}_{ii^{\prime}} (\infty,x) q_{i^{\prime}}(x),
\end{eqnarray}
and the piece from $0$ to $\infty$ is omitted in Eq.~\eqref{eq:wilsonline} since it is irrelevant of LCDA. A  proof is included in Appendix~\ref{identity_wilsonline}.
 
The collinear twist expansion makes use of the decomposition of the quark field into large and small components (see for example \cite{Braun:2003rp})
\begin{eqnarray}
 q= \left(\frac{n\!\!\!\slash \bar n\!\!\!\slash}{2}+ \frac{\bar n\!\!\!\slash  n\!\!\!\slash}{2}\right) q.
\end{eqnarray}
The large component is projected out by $\bar n\!\!\!\slash n\!\!\!\slash$ if quark's flight direction is chosen $\bar n$, i.e.  $p^{\mu}=(p^z,0,0,p^z)$ ($p$ is the momentum of baryon). The twist-3  LCDAs are made of three large  components, and for $\Lambda$ baryon one has the explicit form
\begin{eqnarray}
\Phi(x_1,x_2) f_{\Lambda} u_{\Lambda}(p)=\int  \frac{d t_1p^+}{2\pi} \int \frac{d t_2p^+}{2\pi} e^{ix_1 p^+ t_1 +ix_2 p^+ t_2}\nonumber\\
 \times \epsilon_{ijk}  \langle 0|U_{i}^T(t_1 n) \Gamma D_{j}(t_2 n)  S_{k}(0) |\Lambda\rangle,
\end{eqnarray}
 where $\Gamma=C\gamma^5n\!\!\!\slash$, and $C=i\gamma^2\gamma^0$. $f_{\Lambda}$ is the decay constant for $\Lambda$, and $u_{\Lambda}(p)$ is the $\Lambda$ spinor. The short-distance coefficient is insensitive to the hadrons, i.e. the UV behavior of  LCDAs is irrelevant to the low energy dynamics. In the calculation of LCDAs, one can replace the hadron with a partonic state with the same quantum numbers.   
 
 In the following calculation, we replace the hadron state  $|\Lambda\rangle$ by three constituent quarks state, i.e.  $|\Lambda\rangle \to  | u_{a}(k_1) d_{b}(k_2) s_c(k_3)\rangle$. Here $p=k_1+k_2+k_3$ is the momentum conservation condition. In this case, the leading twist LCDA is defined as
\begin{eqnarray}
\label{eq:Nor_LCDA}
&& \phi(x_1,x_2)S\nonumber\\
 &=& \int  \frac{d t_1p^+}{2\pi} \int \frac{d t_2p^+}{2\pi} e^{ix_1 p^+ t_1 +ix_1 p^+ t_2}  \frac{\epsilon_{ijk}\epsilon_{abc}}{6} 
 \nonumber\\ &\times&  
 \langle 0|U_{i}^T(t_1n) \Gamma D_{j}(t_2n) S_{k}(0) | u_{a}(k_1) d_{b}(k_2) s_c(k_3)\rangle,\nonumber\\
\end{eqnarray} 
where $p^+=n\cdot p$, $x_{i,0}={k_i^+}/{p^+}$, $\sum_{i=1}^3x_i=1$, $\sum_{i=1}^3x_{i,0}=1$, and all longitudinal momentum fractions carried by baryons satisfy $0<x_{i,0}<1$. The normalization factor $S$ can be constructed in terms of the partonic local operator matrix element:
\begin{eqnarray}
\label{eq:local_LCDA}
S&=& \frac{\epsilon_{ijk}\epsilon_{abc}}{6}  
 \langle 0|(U_{i})^T(0) \Gamma D_{j}(0) S_{k}(0) | u_{a}(k_1) d_{b}(k_2) s_c(k_3)\rangle.\nonumber\\
\end{eqnarray}

At tree level, we have
\begin{eqnarray}
\label{eq:local_LCDA-tree}
S^{(0)}&=&-
\frac{\epsilon_{abc}\epsilon_{abc}}{6}  \Big[ {u^{s_1}(k_1)}\Big]^T  \Gamma u^{s_2}(k_2)   u(k_3)
\nonumber \\
&=&2p^+ u(k_3),
\end{eqnarray} 
where $ \Big[ {u^{s_1}(k_1)}\Big]^T  \Gamma u^{s_2}(k_2)=\frac{1}{2}{\rm {tr}}\big[ \slashed {p} C \gamma^5 \Gamma \big]$ is employed, and the superscript index `$(0)$' refer to tree-level result. Here, the quark state is chosen to have the same $J^{PC}$ with the $\Lambda$, and the spin average and color average are assumed in this calculation. In Appendix~\ref{App:trace_formulae}, we provide a detailed explanation of Eq.~\eqref{eq:local_LCDA-tree}, and the corresponding trace formalism to derive this convention. 

After a bit of algebra, we obtain the result of LCDA at the tree level 
\begin{eqnarray}
\phi^{(0)}(x_1,x_2,\mu)S^{(0)}&=&
 \int  \frac{d t_1p^+}{2\pi} \int \frac{d t_2p^+}{2\pi} e^{ip^+(x_1 t_1+x_2t_2)}\nonumber\\&& \times\Big[ {u^{s_1}(k_1)}\Big]^T  \Gamma u^{s_2}(k_2) u(k_s)\nonumber\\&& \times e^{-ik_1^+ t_1}  e^{-ik_2^+ t_2}\nonumber\\
&=&\delta(x_1-x_{1,0})\delta(x_2-x_{2,0})S^{(0)},
\end{eqnarray}
i.e. 
\begin{eqnarray}
\phi^{(0)}(x_1,x_2,\mu)
&=&\delta(x_1-x_{1,0})\delta(x_2-x_{2,0}).
\end{eqnarray}

\begin{figure}[!t]
\includegraphics[width=0.45\textwidth]{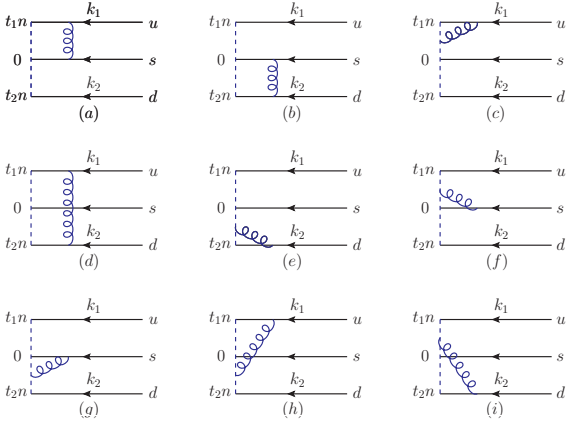} 
\caption{One-loop corrections to LCDAs of a light baryon $\Lambda$.}
\label{fig:DAdiagram}
\end{figure}

At the one-loop order, two gluons are radiated from (1)  light quarks by the QCD interactions and (2) two pieces of gauge-link. These five objects give $C_5^2=10$ terms with five self-energy corrections in total. The diagram from two gauge links is zero since $n^2=0$ and the rest is displayed in Fig.~\ref{fig:DAdiagram} (quark self-energy corrections are not shown). We choose the
dimensional regularization $D = 4-2\epsilon$ to regularize the UV and IR divergences.

The real diagram shown in Fig.~\ref{fig:DAdiagram}(a) can be obtained:
\begin{eqnarray}
\label{eq:LightConeOnShellPhia}
 \mathcal{M}^{a} &=& ig^2\frac{-C_F}{2}(p^+)^2\Big(\frac{\mu^2}{e^{\ln(4\pi)-\gamma_E}}\Big)^{\epsilon} \int \frac{d^D q}{(2\pi)^D} \frac{1}{q^2+i\epsilon}\nonumber \\&&\times \frac{1 }{(k_3-q)^2+i\epsilon} \frac{\delta(x_1p^+-q^+-k_1^+)\delta(x_2p^+-k_2^+)}{(q+k_1)^2+i\epsilon} \nonumber \\&&\times  \big[{u^{s_1}(k_1)}\big]^T\gamma^{\mu}(\slashed{q}+{\slashed{k_1}})\Gamma u^{s_2}(k_2) ({\slashed{k_3}}-{\slashed{q}})\gamma_{\mu}u(k_3)
 \nonumber\\
 &=& 
 ig^2\frac{-C_F}{2}(p^+)^2\Big(\frac{\mu^2}{e^{\ln(4\pi)-\gamma_E}}\Big)^{\epsilon} \int \frac{d^D q}{(2\pi)^D} \frac{1}{q^2+i\epsilon} \nonumber\\&&\times\frac{\delta(x_1p^+-q^+-k_1^+)\delta(x_2p^+-k_2^+)}{(q+k_1)^2+i\epsilon} \frac{1 }{(k_3-q)^2+i\epsilon}\
  \nonumber \\&& \times \frac{1}{2}(-1){\rm tr}\left(\slashed {p} \gamma^5 \gamma^{\mu} (\slashed{q}+\slashed{k_1})\gamma^5 \slashed{n}\right)({\slashed{k_3}}-{\slashed{q}})\gamma_{\mu}u(k_3),\nonumber\\
\end{eqnarray}
where $\mu$ is the renormalization scale in the $\overline {\rm{MS}}$ scheme,  and $\mathcal{M}$ denotes $\phi(x_1,x_2,\mu)S$ in short. The color factor is $-\frac{C_F}{2}$ which comes from color Fierz transformation:
\begin{eqnarray}
\frac{\epsilon_{ijk}\epsilon_{abc}}{6}T^A_{ia}T^A_{kc}\delta_{jb}=\frac{\epsilon_{ijk}\epsilon_{abc}}{6}\frac{\delta_{jb}}{2}\left(\delta_{ic}\delta_{ak}-\frac{1}{N_c}\delta_{ia}\delta_{kc}\right).\nonumber\\
\end{eqnarray}

The spinor part in the last line of Eq.~\eqref{eq:LightConeOnShellPhia} can be projected out by taking the projection technique, i.e. 
\begin{eqnarray}
&&\frac{1}{2}\times(-1){\rm tr}\left(\slashed {p} \gamma^5 \gamma^{\mu} (\slashed{q}+\slashed{k_1})\gamma^5 \slashed{n}\right)({\slashed{k_3}}-{\slashed{q}})\gamma_{\mu}u(k_3)\nonumber\\&=&\big(-q_{\perp}^2-q^-q_{\perp}\slashed{n}\slashed{n}_{\perp}\big)S^{(0)},
\end{eqnarray}
and the integration of the second term is zero because the integrated function for $q_\perp$ is odd. Finally, the amplitude of Fig.~\ref{fig:DAdiagram}(a) can be simplified to
\begin{eqnarray}
&&\mathcal{M}^{a}=\frac{\alpha_s C_F}{8\pi} \delta(x_2-x_{2,0}) \theta(x_1)\theta(x_3) \theta(1-x_1)\theta(1-x_3) \nonumber\\&& \times \Bigg[\frac{x_3\theta(x_1-x_{1,0})}{(1-x_2) x_{3,0}}+\frac{x_1\theta(x_{1,0}-x_1)}{(1-x_2) x_{1,0}}\Bigg]\Big(\frac{1}{\epsilon_{\rm UV}}-\frac{1}{\epsilon_{\rm IR}}\Big)S^{(0)},\nonumber\\
\end{eqnarray}
where we use $\alpha_s=\alpha_s(\mu)$ in short.

In a similar way, the real diagram Fig.~\ref{fig:DAdiagram}(d) can be obtained as follows:
\begin{eqnarray}
\label{eq:LightConeOnShellPhid}
 \mathcal{M}^{d}&=&ig^2\frac{-C_F}{2}(p^+)^2  \Big(\frac{\mu^2}{e^{\ln(4\pi)-\gamma_E}}\Big)^{\epsilon}  \int \frac{d^D q}{(2\pi)^D}\frac{1}{q^2+i\epsilon}   \nonumber \\&& \times 
\frac{1}{(k_2-q)^2+i\epsilon}   \frac{ \delta(\omega_1-k_1^+-q^+)\delta(\omega_2-k_2^++q^+)  }{(k_1+q)^2+i\epsilon}      \nonumber \\&&\times 
\big[{u^{s_1}(k_1)}\big]^T\gamma^{\mu}(\slashed{k_1}+\slashed{q})\Gamma(\slashed{k_2}-\slashed{q})\gamma_{\mu}u^{s_2}(k_2)u(k_3) \nonumber \\
&=&\frac{\alpha_s C_F}{4\pi}\delta(x_3-x_{3,0}) \theta(x_1)\theta(x_2)\theta(1-x_1)\theta(1-x_2)\nonumber\\&&\times
 \Big[\frac{x_2\theta(x_1-x_{1,0})}{(x_1+x_2)x_{2,0}} +\frac{x_1\theta(x_2-x_{2,0})}{(x_1+x_2)x_{1,0}}\Big](\frac{1}{\epsilon_{\rm UV}}-\frac{1}{\epsilon_{\rm IR}})S^{(0)}.\nonumber\\
\end{eqnarray}
This result is symmetric under the exchange $x_1 \leftrightarrow x_2$ and $x_{1,0} \leftrightarrow x_{2,0}$. At the same time, we should also note that in addition to the color factor $-\frac{C_F}{2}$, the normalization factor $S^{(0)}$, and the fraction $\frac{\delta(x_3-x_{3,0})}{x_1+x_2}$, which is the same as the contribution of the pseudo scalar $\pi$ meson distribution amplitude to the external leg exchange gluon diagram.

The result of Fig.~\ref{fig:DAdiagram}(b) can be obtained from the result of Fig.~\ref{fig:DAdiagram}(a) with $x_2 \leftrightarrow x_1$ and $x_{2,0} \leftrightarrow x_{1,0}$. Therefore, we can write the result of Fig.~\ref{fig:DAdiagram}(b) as follows
\begin{eqnarray}
&&\mathcal{M}^{b}=\frac{\alpha_s C_F}{8\pi} \delta(x_1-x_{1,0}) \theta(x_2)\theta(x_3)\theta(1-x_2)\theta(1-x_3) \nonumber\\&&\times\Bigg[\frac{x_3\theta(x_2-x_{2,0})}{(1-x_1) x_{3,0}}+\frac{x_2\theta(x_{2,0}-x_2)}{(1-x_1) x_{2,0}}\Bigg]\Big(\frac{1}{\epsilon_{\rm UV}}-\frac{1}{\epsilon_{\rm IR}}\Big)S^{(0)}. \nonumber\\
\end{eqnarray}

For the diagram Fig.~\ref{fig:DAdiagram}(c), we have
\begin{eqnarray}
\label{eq:LightConeOnShellPhic}
 &\mathcal{M}^{c} 
 &= -ig^2C_F p^+\delta(x_2-x_{2,0})\Big(\frac{\mu^2}{e^{\ln(4\pi)-\gamma_E}}\Big)^{\epsilon} \int \frac{d^D q}{(2\pi)^D}\frac{1}{q^+} \nonumber \\&&\times \frac{1}{q^2+i\epsilon} \frac{\delta\big((x_1-x_{1,0})p^+-q^+\big)-\delta\big((x_1-x_{1,0})p^+\big)}{(q+k_1)^2+i\epsilon} \nonumber \\&&
\times (u^{s_1}(k_1))^T\gamma^{\mu}(\slashed{q}+\slashed{k_1}) \Gamma u^{s_2}(k_2)n_{\mu}u(k_3).
\end{eqnarray}
We should note that the color factor $C_F$ in diagram Fig.~\ref{fig:DAdiagram}(c) and its symmetric diagram  Fig.~\ref{fig:DAdiagram}(e) are different from the other graphs. That is because Fig.~\ref{fig:DAdiagram}(c) and Fig.~\ref{fig:DAdiagram}(e) have no change the color structure. After simplifying Eq.~\eqref{eq:LightConeOnShellPhic}, we have
\begin{eqnarray}
\mathcal{M}^{c}&=&-\frac{\alpha_s C_F}{2\pi}  \Bigg[\delta(x_2-x_{2,0})  \theta(x_1) \theta(x_3) \theta(1-x_1) \theta(1-x_3) 
\nonumber \\&&\times \frac{\theta(x_{1,0}-x_1)}{(x_1-x_{1,0})} \frac{x_1 }{x_{1,0}}\Bigg]_\oplus \big(\frac{1}{\epsilon_{\rm UV}}-\frac{1}{\epsilon_{\rm IR}}\big)S^{(0)},
\end{eqnarray}
where the $\oplus$ denote
\begin{eqnarray}
\left[g\left(x_1, x_2, x_{1,0}, x_{2,0}\right)\right]_{\oplus}&=&g\left(x_1, x_2, x_{1,0}, x_{2,0}\right)\nonumber\\&&
-\delta\left(x_1-x_{1,0}\right)\delta\left(x_2-x_{2,0}\right)\nonumber\\
&&\times \int d y_1d y_2 g\left(y_1, y_2, x_{1,0}, x_{2,0}\right).\nonumber\\
\end{eqnarray}

The result of Fig.~\ref{fig:DAdiagram}(e) can be also obtained from the result of Fig.~\ref{fig:DAdiagram}(c) with $x_2 \leftrightarrow x_1$ and $x_{2,0} \leftrightarrow x_{1,0}$,
\begin{eqnarray}
\label{eq:LightConeOnShellPhie}
\mathcal{M}^e&=&-\frac{\alpha_s C_F}{2\pi} \Bigg[\delta(x_1-x_{1,0})  \theta(x_2) \theta(x_3) \theta(1-x_2) \theta(1-x_3) 
\nonumber \\&&\times  \frac{\theta(x_{2,0}-x_2)}{(x_2-x_{2,0})} \frac{x_2 }{x_{2,0}}\Bigg]_\oplus \big(\frac{1}{\epsilon_{\rm UV}}-\frac{1}{\epsilon_{\rm IR}}\big)S^{(0)}.
\end{eqnarray}
We should also notice that the results for Fig.~\ref{fig:DAdiagram}(c) and Fig.~\ref{fig:DAdiagram}(e) are the similar to the result of the $\pi$ meson distribution amplitude to the diagram which connected the quark and the Wilson line.

For completeness of the calculation, we present the results of the other two diagrams Fig.~\ref{fig:DAdiagram}(f)(h).
\begin{eqnarray}
\label{eq:LightConeOnShellPhif}
 \mathcal{M}^{f}&=&-ig^2 \frac{-C_F}{2}
p^+\delta(x_2-x_{2,0})\Big(\frac{\mu^2}{e^{\ln(4\pi)-\gamma_E}}\Big)^{\epsilon} \int \frac{d^D q}{(2\pi)^D}\nonumber\\&&\frac{\delta(x_1p^+-k_1^+-q^+)-\delta(x_1p^+-k_1^+)}{q^+(q^2+i\epsilon)}\frac{1}{(k_3-q)^2+i\epsilon}\nonumber \\&&\times \big[{u^{s_1}(k_1)}\big]^T\Gamma u^{s_2}(k_2) n^{\mu}(\slashed{k_3}-\slashed{q})\gamma_{\mu}u(k_3),\nonumber \\
&=&\frac{\alpha_s C_F}{4\pi}
\Bigg[\delta(x_2-x_{2,0})\theta(x_1)\theta(1-x_1)\theta(x_3)\theta(1-x_3)\nonumber\\&&\times\frac{\theta(x_1-x_{1,0})x_3}{(x_1-x_{1,0})x_{3,0}} \Bigg]_\oplus \big(\frac{1}{\epsilon_{\rm UV}}-\frac{1}{\epsilon_{\rm IR}}\big)S^{(0)},
\end{eqnarray}
and
\begin{eqnarray}
\label{eq:LightConeOnShellPhih}
 \mathcal{M}^{h}&=&-ig^2 \frac{-C_F}{2}
(p^+)^2\Big(\frac{\mu^2}{e^{\ln(4\pi)-\gamma_E}}\Big)^{\epsilon} \int \frac{d^D q}{(2\pi)^D}\nonumber\\&&
\times \frac{\delta(x_2p^+-k_2^+)-\delta(x_2p^++q^+-k_2^+)}{q^+} \nonumber \\&& \times \frac{\delta(x_1p^+-q^+-k_1^+)}{\big[q^2+i\epsilon\big]\big[(q+k_1)^2+i\epsilon\big]} \nonumber\\ &&\times \big[u^{s_1}(k_1)\big]^T\gamma^{\mu}(\slashed{q}+\slashed{k_1})
\Gamma n_{\mu} (u^{s_2}(k_2))u(k_3),\nonumber\\
&=&\frac{\alpha_s C_F}{4\pi}
\Big[\delta(x_2-x_{2,0})-\delta(x_3-x_{3,0})\Big]\theta(x_1)\theta(1-x_1)\nonumber\\&&\times
\frac{x_1\theta(x_{1,0}-x_1)}{(x_1-x_{1,0})x_{1,0}} \Big(\frac{1}{\epsilon_{\rm UV}}-\frac{1}{\epsilon_{\rm IR}}\Big)S^{(0)}.
\end{eqnarray}
Therefore,
\begin{eqnarray}
\label{eq:LightConeOnShellPhig}\mathcal{M}^{g}&=&\frac{\alpha_s C_F}{4\pi}
\Bigg[\delta(x_1-x_{1,0})\theta(x_2)\theta(1-x_2)\theta(x_3)\theta(1-x_3)\nonumber\\&&\times\frac{\theta(x_2-x_{2,0})x_3}{(x_2-x_{2,0})x_{3,0}} \Bigg]_\oplus \big(\frac{1}{\epsilon_{\rm UV}}-\frac{1}{\epsilon_{\rm IR}}\big)S^{(0)},
\end{eqnarray}
and
\begin{eqnarray}
\label{eq:LightConeOnShellPhii}
 \mathcal{M}^{i}&=&\frac{\alpha_s C_F}{4\pi}
\Big[\delta(x_1-x_{1,0})-\delta(x_3-x_{3,0})\Big]\theta(x_2)\theta(1-x_2)\nonumber\\&&\times
\frac{x_2\theta(x_{2,0}-x_2)}{(x_2-x_{2,0})x_{2,0}} \Big(\frac{1}{\epsilon_{\rm UV}}-\frac{1}{\epsilon_{\rm IR}}\Big)S^{(0)}.
\end{eqnarray}

Combining the above results with Eq.~\eqref{app:nor_LCDA}, we have the complete result for the one-loop normalized and $\overline{\rm MS}$ renormalized LCDA  as 
\begin{eqnarray}
\label{eq:oneloop_Nor_LCDA}
\phi(x_1,x_2,\mu)&=&\delta(x_1-x_{1,0})\delta(x_2-x_{2,0})\nonumber\\&&-\frac{\alpha _s C_F }{8 \pi }\frac{1}{\epsilon_{\rm IR}}f(x_1,x_2,x_{1,0},x_{2,0}),
\end{eqnarray}
with
\begin{eqnarray}
&&f(x_1,x_2,x_{1,0},x_{2,0})=\nonumber\\&&
\Bigg\{\frac{\delta(x_1-x_{1,0})\theta(x_1)\theta(\bar{x}_1)\theta(x_2)\theta(\bar{x}_2)}{\bar{x}_1\left(x_2-x_{2,0}\right)}\nonumber\\&& \times
\Big[\frac{x_3 \left(x_2-x_{2,0}+2\bar{x}_1\right) }{x_{3,0}}\theta \left(x_2-x_{2,0}\right)\nonumber\\&&+\frac{x_2 \left(x_2-x_{2,0}-2\bar{x}_1\right) }{x_{2,0}}\theta \left(x_{2,0}-x_2\right)\Big]\nonumber\\&&+
\frac{\delta(x_3-x_{3,0})2 x_1\theta(x_1)\theta(\bar{x}_1)\theta(x_2)\theta(\bar{x}_2) }{x_{1,0}} \nonumber\\&& \times \Big[\frac{\theta \left(x_2-x_{2,0}\right)}{x_1+x_2}-\frac{\theta \left(x_{1,0}-x_1\right)}{x_1-x_{1,0}}\Big]\nonumber\\&&+\{ x_1 \leftrightarrow x_2,x_{1,0} \leftrightarrow x_{2,0}\}\Bigg\}_\oplus ,
\end{eqnarray}
where $\bar{x}=1-x$.

The renormalized LCDA can be obtained by removing the UV divergence due to   the renormalization of the composite operator of LCDA. The dependence of the renormalized LCDA $\Phi(x_1,x_2,\mu)$ on $\ln \mu$ can be obtained from the evolution equation
\begin{eqnarray}
\label{eq:evolution}
\frac{d \Phi(x_1,x_2,\mu)}{d \ln \mu}=\frac{\alpha_s C_F}{4\pi}\int d y_1 \int d y_2 V\Phi(y_1,y_2,\mu).  
\end{eqnarray}
At the one-loop, the evolution kernel $V$ is
\begin{eqnarray}
V=f(x_1,x_2,y_1,y_2).
\end{eqnarray}
We should also note that the form of Eq.~\eqref{eq:evolution} is similar to the  Efremov-Radyushkin-Brodsky-Lepage (ERBL) evolution equation for mesons~\cite{Lepage:1979zb,Efremov:1979qk}.
   
\section{Quasi-DA at  one loop level}\label{sec:Quasi-DA}

In this section, we will introduce an equal-time operator matrix element which is often named as quasi-distribution amplitudes \cite{Ji:2013dva}. The quasi-DA for the $\Lambda$  is defined as
\begin{eqnarray}
&\tilde{\Phi}(x_1,x_2,\mu)\tilde{f}_{\Lambda} u_{\Lambda}(p)&=\int  \frac{d t_1p^z}{2\pi} \int \frac{d t_2p^z}{2\pi} e^{ix_1 p^z t_1 +ix_1 p^z t_2}\nonumber\\&&
 \times \epsilon_{ijk}  \langle 0|U_{i}^T(t_1 n_z) \tilde{\Gamma} D_{j}(t_2 n_z)  S_{k}(0) |\Lambda\rangle,\nonumber\\
\end{eqnarray}
where $\tilde{f}_{\Lambda}$ is the quasi decay constant for $\Lambda$.

In a similar way, the corresponding partonic operator matrix element is
\begin{eqnarray}
\label{eq:Nor_quasiLCDA}
&& \tilde{\phi}(x_1,x_2,\mu)\tilde{S}\nonumber\\
 &=& \int  \frac{d t_1p^z}{2\pi} \int \frac{d t_2p^z}{2\pi} e^{ix_1 p^z t_1 +ix_1 p^z t_2}  \frac{\epsilon_{ijk}\epsilon_{abc}}{6}\times \nonumber\\ &&  
 \langle 0|(U_{i})^T(t_1n_z) \tilde{\Gamma} D_{j}(t_2n_z) S_{k}(0) | u_{a}(k_1) d_{b}(k_2) s_c(k_3)\rangle.\nonumber\\
\end{eqnarray} 
The $\tilde{\Gamma}$ is the Dirac matrix for the quasi-DA and two popular choices are $\tilde{\Gamma}=C\gamma^5\slashed{n}_{\lambda}$ ($\lambda=t$ or $z$) for quasi-DA. The two choices will give the same results at leading twist and a brief explanation is given in  Appendix~\ref{App:Expansion_by_region}. Here $n_t^{\mu} = (1,0,0,0)$ and $n_z^{\mu} = (0,0,0,-1)$. The corresponding normalization factor $\tilde{S}$ is 
\begin{eqnarray}
\label{eq:local_quasiLCDA}
\tilde{S}&=& \frac{\epsilon_{ijk}\epsilon_{abc}}{6}  \langle 0|(U_{i})^T(0) \tilde{\Gamma} D_{j}(0) S_{k}(0) | u_{a}(k_1) d_{b}(k_2) s_c(k_3)\rangle.\nonumber\\
\end{eqnarray}

At the tree level, we have the matrix element
\begin{eqnarray}
\tilde{S}^{(0)}&=&-
\frac{\epsilon_{abc}\epsilon_{abc}}{6}  \Big[ {u^{s_1}(k_1)}\Big]^T  \tilde{\Gamma} u^{s_2}(k_2)   u(k_3)
\nonumber \\
&=&2p^z u(k_3),
\end{eqnarray} 
where $ \Big[ {u^{s_1}(k_1)}\Big]^T  \tilde{\Gamma} u^{s_2}(k_2)=\frac{1}{2}{\rm {tr}}\big[ \slashed {p} C \gamma^5 \tilde{\Gamma} \big]$ is employed.
The tree-level result for quasi-DA is
\begin{eqnarray}
\tilde{\phi} ^{(0)}(x_1,x_2,\mu)\tilde{S}^{(0)}&=&{p^z}^2
\int \frac{d t_1}{2\pi} \int \frac{d t_2}{2\pi} e^{ix_1 t_1 p^z+ix_2 t_2 p^z} \nonumber\\&& \times \Big[ {u^{s_1}(k_1)}\Big]^T  \tilde{\Gamma} u^{s_2}(k_2) u(k_3)\nonumber\\&& \times e^{-ix_{1,0}p^z t_1} e^{-ix_{2,0}p^z t_2}\nonumber\\
&=&\delta(x_1-x_{1,0})\delta(x_2-x_{2,0}) \tilde{S}^{(0)}.
\end{eqnarray}
Therefore, the normalized quasi-DA at the tree level is
\begin{eqnarray}
\tilde{\phi}^{(0)}(x_1,x_2,\mu)=\delta(x_1-x_{1,0})\delta(x_2-x_{2,0}).
\end{eqnarray}
We can find that the normalized LCDA  at the tree-level gives the same result as the quasi-DA. The one-loop diagrams of quasi-DA for baryon $\Lambda$ are similar to that of the LCDA which are shown in Fig.~\ref{fig:DAdiagram}, except the $n$ direction is changed to $n_z$.  The real diagram for quasi-DA shown in Fig.~\ref{fig:DAdiagram}(a) can be obtained as follows:
\begin{eqnarray}\label{eq:quasifiga}
 \tilde{\mathcal{M}}^{a} 
 &=& ig^2\frac{-C_F}{2} {p^z}^2 \Big(\frac{\mu^2}{e^{\ln(4\pi)-\gamma_E}}\Big)^{\epsilon} \int \frac{d^D q}{(2\pi)^D} \frac{1}{q^2+i\epsilon}\nonumber\\&&\times\frac{1}{(k_3-q)^2+i\epsilon} \frac{\delta(x_1p^z-q^z-{k_1}^z)\delta(x_2p^z-{k_2}^z)}{(q+k_1)^2+i\epsilon}  \nonumber \\&&\times\big[{u^{s_1}(k_1)}\big]^T\gamma^{\mu}(\slashed{q}+{\slashed{k_1}})\tilde{\Gamma} u^{s_2}(k_2) ({\slashed{k_3}}-{\slashed{q}})\gamma_{\mu}u(k_3),
 \nonumber\\&=& ig^2\frac{-C_F}{2} {p^z}^2 \Big(\frac{\mu^2}{e^{\ln(4\pi)-\gamma_E}}\Big)^{\epsilon} \int \frac{d^D q}{(2\pi)^D} \frac{1}{q^2+i\epsilon}\nonumber\\&&\times\frac{1}{(k_3-q)^2+i\epsilon} \frac{\delta(x_1p^z-q^z-{k_1}^z)\delta(x_2p^z-{k_2}^z)}{(q+k_1)^2+i\epsilon}  \nonumber \\&&\times
(-\frac{1}{2}){\rm tr}\left(\slashed {p} \gamma^5 \gamma^{\mu} (\slashed{q}+\slashed{k_1})\gamma^5 \slashed{n}_{\lambda}\right)({\slashed{k_3}}-{\slashed{q}})\gamma_{\mu}u(k_3),\nonumber\\
 \end{eqnarray}
where $\tilde{\mathcal{M}}$ denotes $\tilde{\phi}(x_1,x_2,\mu)\tilde{S}$ in short. The last line  in the above equation reads
\begin{eqnarray}
&&-\frac{1}{2}{\rm tr}\left(\slashed {p} \gamma^5 \gamma^{\mu} (\slashed{q}+\slashed{k_1})\gamma^5 \gamma^{\lambda}\right)({\slashed{k_3}}-{\slashed{q}})\gamma_{\mu}u(k_3)
\nonumber\\
&=&\big[q^2-(p\cdot q)\frac{q^t+q^z}{p^z}-q_{\perp}\slashed{n}_{\perp}\slashed{n}_z\big]\tilde{S}^{(0)},
\end{eqnarray} 
and the third term gives zero contribution because the integrand as in Eq.~\eqref{eq:quasifiga}  is odd in  $q_\perp$. This result indicates the equivalence of the two Lorentz structures $\tilde{\Gamma}$.

Finally, the quasi-DA in Fig.~\ref{fig:DAdiagram}(a) can be simplified as
\begin{widetext}
\begin{eqnarray}
&&\tilde{\mathcal{M}}^{a}
 =\frac{\alpha_s C_F}{8\pi} \tilde{S}^{(0)} \delta(x_2-x_{2,0})\nonumber \\&& \times 
\left\{ \begin{aligned}
&\Big[\frac{x_1  \ln \frac{-x_1}{x_3}}{(1-x_{2,0})x_{3,0}}-\frac{x_1\ln \frac {-x_1}{x_{1,0}-x_1}}{x_{1,0}x_{3,0}}-\frac{
\ln \frac{x_{1,0}-x_1}{x_3}}{x_{3,0}}\Big],~x_1<0\\
&\Big[\frac{x_1 \left(\ln \frac{(x_{1,0}-x_1)x_1}{\mu^2/({2p^z})^2}-1\right)}{x_{1,0} x_{3,0}}-\frac{x_1 \left(\ln \frac{x_1x_3}{\mu^2/({2p^z})^2} -1\right)}{\left(1-x_{2,0}\right) x_{3,0}}
-\frac{ \ln \frac{x_{1,0}-x_1}{x_3} }{ x_{3,0}}-\frac{x_1}{x_{1,0}\left(1-x_{2,0}\right)}\frac{1}{\epsilon_{\rm IR}}\Big],~0<x_1<x_{1,0}\\
&\Big[\frac{x_1 \ln \frac{x_1}{x_1-x_{1,0} }}{x_{1,0} x_{3,0}}-\frac{x_1 \left(\ln \frac{x_1x_3}{\mu^2/({2p^z})^2} -1\right)}{\left(1-x_{2,0}\right) x_{3,0}}
+\frac{ \ln \frac{(x_1-x_{1,0})x_3}{ \mu^2/({2p^z})^2}-1}{ x_{3,0}}-\frac{x_3}{x_{3,0}\left(1-x_{2,0}\right)}\frac{1}{\epsilon_{\rm IR}}\Big],~x_{1,0}<x_1<1-x_2\\
&\Big[\frac{x_1 \ln \frac{x_1}{x_1-x_{1,0} }}{x_{1,0} x_{3,0}}+\frac{x_1 \ln \frac{-x_3}{x_1} }{\left(1-x_{2,0}\right) x_{3,0}}+\frac{ \ln \frac{x_1-x_{1,0}}{-x_3}}{ x_{3,0}}\Big]. ~x_1>1-x_2\\
\end{aligned}\right. 
\end{eqnarray}
\end{widetext}
For the remaining diagrams, rather than enumerating the calculations in detail, we directly give their  results
\begin{eqnarray}
\tilde{\mathcal{M}}^{b} =\tilde{\mathcal{M}}^{a} |_{x_2 \leftrightarrow x_1, x_{2,0} \leftrightarrow x_{1,0}},
\end{eqnarray}
\begin{widetext}

\begin{eqnarray}
&&\tilde{\mathcal{M}}^{d} =\frac{\alpha_s C_F}{4\pi} \tilde{S}^{(0)}\delta(x_1+x_2-x_{1,0}-x_{2,0}) \nonumber \\&&
\times \left\{ \begin{aligned}
&\Big[\frac{x_1}{x_{1,0}(x_{1,0}+x_{2,0})}\ln\frac{x_2}{-x_1}+\frac{x_{2,0}
   -x_2}{x_{1,0} x_{2,0}}\ln\frac{x_2-x_{2,0}}{x_2}\Big],~x_1 < 0\\
&\Big\{\frac{x_1 \Big[\ln \left(\frac{x_1x_2}{\mu^2/({2p^z})^2}\right)+1\Big]}{x_{1,0}\left(x_{1,0}+x_{2,0}\right)}
+\frac{\left(x_2-x_{2,0}\right) \ln \frac{x_2}{x_2-x_{2,0}}}{x_{1,0}x_{2,0}}-\frac{x_1}{x_{1,0}\left(x_{1,0}+x_{2,0}\right)} \frac{1}{\epsilon_{\rm IR}}\Big\},~0<x_1 <x_{1,0}\\
&\Big\{\frac{x_1 \Big[\ln \left(\frac{x_1x_2}{\mu^2/({2p^z})^2}\right)+1\Big]}{x_{1,0}\left(x_{1,0}+x_{2,0}\right)}
+\frac{\left(x_2-x_{2,0}\right)\Big[\ln \frac{x_2(x_{2,0}-x_2)}{\mu^2/({2p^z})^2}+1\Big]}{x_{1,0}x_{2,0}}-\frac{x_2}{x_{2,0}\left(x_{1,0}+x_{2,0}\right)}\frac{1}{\epsilon_{\rm IR}}\Big\},x_{1,0}<x_1 <x_1+x_2\\
&\Big[\frac{x_1}{x_{1,0}(x_{1,0}+x_{2,0})}\ln \frac{x_1}{-x_2}+\frac{x_{2,0}
   -x_2 }{x_{1,0}x_{2,0}}\ln \frac{-x_2}{x_{2,0}-x_2}\Big],~x_1 > x_1+x_2
\end{aligned}\right. 
\end{eqnarray}

\begin{eqnarray}
&&\tilde{\mathcal{M}}^{c} =\frac{\alpha_s C_F}{4\pi} \tilde{S}^{(0)}\nonumber \\&&
\times 
\left\{ \begin{aligned}
&\Bigg\{\delta(x_2-x_{2,0})\Big[ \frac{1}{x_{1,0}-x_1}-\frac{2x_1\ln \frac{-x_1}{x_{1,0}-x_1}}{x_{1,0}(x_{1,0}-x_1)} \Big]\Bigg\}_\oplus,~x_1<0\\
&\Bigg\{\delta(x_2-x_{2,0})\Big[ \frac{1}{x_{1,0}-x_1}-\frac{2x_1\Big[\ln \frac{(x_{1,0}-x_1)x_1}{\mu^2/({2p^z})^2}-1\Big]}{x_{1,0}(x_1-x_{1,0})}+\frac{2x_1}{x_{1,0}(x_1-x_{1,0})}\frac{1}{\epsilon_{\rm IR}}\Big]\Bigg\}_\oplus,~0<x_1<x_{1,0}\\
&\Bigg\{\delta(x_2-x_{2,0})\Big[ \frac{1}{x_1-x_{1,0}}-\frac{2x_1\ln \frac{x_1}{x_1-x_{1,0}}}{x_{1,0}(x_1-x_{1,0})}\Big]\Bigg\}_\oplus,~x_1>x_{1,0}\\
\end{aligned}\right.
\end{eqnarray}

\begin{eqnarray}
\tilde{\mathcal{M}}^{e} =\tilde{\mathcal{M}}^{c} |_{x_2 \leftrightarrow x_1, x_{2,0} \leftrightarrow x_{1,0}}
\end{eqnarray}

\begin{eqnarray}
&&\tilde{\mathcal{M}}^{f} =\frac{\alpha_s C_F}{8\pi} \tilde{S}^{(0)}\nonumber \\&&
\times 
\left\{ \begin{aligned}
&\Bigg\{\delta(x_2-x_{2,0})\Big[ \frac{1}{x_{1,0}-x_1}+\frac{2x_3\ln \frac{x_{1,0}-x_1}{x_3 }}{x_{3,0}(x_{1,0}-x_1)}  \Big]\Bigg\}_\oplus,~x_1<x_{1,0}\\
&\Bigg\{\delta(x_2-x_{2,0})\Big[\frac{1}{x_1-x_{1,0}}+\frac{2x_3\Big[\ln \frac{(x_1-x_{1,0})x_3}{\mu^2/({2p^z})^2}-1\Big] }{x_{3,0}(x_1-x_{1,0})}
-\frac{2x_3}{x_{3,0}(x_1-x_{1,0})}\frac{1}{\epsilon_{\rm IR}}
\Big]\Bigg\}_\oplus,~x_{1,0}<x_1<1-x_2\\
&\Bigg\{\delta(x_2-x_{2,0})\Big[ \frac{1}{x_1-x_{1,0}}+\frac{2x_3\ln \frac{x_1-x_{1,0}}{-x_3 }}{x_{3,0}(x_1-x_{1,0})} \Big]\Bigg\}_\oplus,~x_1>1-x_2
\end{aligned}\right.
\end{eqnarray}

\begin{eqnarray}
\tilde{\mathcal{M}}^{g} =\tilde{\mathcal{M}}^{f} |_{x_2 \leftrightarrow x_1, x_{2,0} \leftrightarrow x_{1,0}}
\end{eqnarray}

 \begin{eqnarray}
&&\tilde{\mathcal{M}}^{h} =\frac{\alpha_s C_F}{8\pi} \tilde{S}^{(0)} \Big[\delta(x_3-x_{3,0})-\delta(x_2-x_{2,0})\Big]\nonumber\\&&\times
\left\{ \begin{aligned}
&\Big[ \frac{1}{x_{1,0}-x_1}+\frac{2x_1\ln \frac{x_{1,0}-x_1}{-x_1 }}{x_{1,0}(x_{1,0}-x_1)}  \Big],~x_1<0\\
&\Big\{\frac{1}{x_{1,0}-x_1}-\frac{2x_1\Big[\ln \frac{(x_{1,0}-x_1)x_1}{\mu^2/({2p^z})^2}-1\Big] }{x_{1,0}(x_1-x_{1,0})}
+\frac{2x_1}{x_{1,0}(x_1-x_{1,0})}\frac{1}{\epsilon_{\rm IR}}
\Big\},~0<x_1<x_{1,0}\\
&\Big[ \frac{1}{x_1-x_{1,0}}+\frac{2x_1\ln \frac{x_1-x_{1,0}}{x_1}}{x_{1,0}(x_1-x_{1,0})} \Big],~x_1>x_{1,0}
\end{aligned}\right.
\end{eqnarray}

\begin{eqnarray}
\tilde{\mathcal{M}}^{i} =\tilde{\phi}^{h} |_{x_2 \leftrightarrow x_1, x_{2,0} \leftrightarrow x_{1,0}}
\end{eqnarray}
\end{widetext}

\begin{figure}[!t]
\includegraphics[width=0.45\textwidth]{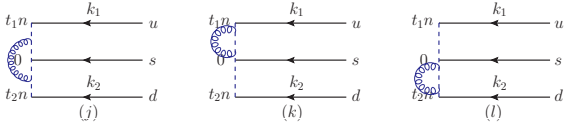} 
\caption{One loop self-energy corrections for quasi-DA.}
\label{fig:Wilson-line_self-energy}
\end{figure}

In addition, since $n_z^2 \neq 0$, the self-energy diagram of the Wilson line of the quasi-DA also contributes. These two Wilson lines give three terms of one-loop self-energy corrections which are shown in Fig.~\ref{fig:Wilson-line_self-energy}. Those three self-energy reads

\begin{eqnarray}
\tilde{\mathcal{M}}^{j}&=&\frac{\alpha_s C_F}{4\pi}\tilde{S}^{(0)}\Bigg[\frac{\delta(x_2-x_{2,0})}{\sqrt{(x_1-x_{1,0})^2}}+\frac{\delta(x_1-x_{1,0})}{\sqrt{(x_2-x_{2,0})^2}}\nonumber\\&&-\frac{\delta(x_3-x_{3,0})}{\sqrt{(x_1-x_{1,0})^2}}\Bigg]_\oplus ,\nonumber\\
\tilde{\mathcal{M}}^{k}&=&-\frac{\alpha_s C_F}{2\pi}\tilde{S}^{(0)}\Big[\frac{\delta(x_2-x_{2,0})}{\sqrt{(x_1-x_{1,0})^2}}\Big]_\oplus,\nonumber\\
\tilde{\mathcal{M}}^{l}&=&-\frac{\alpha_s C_F}{2\pi}\tilde{S}^{(0)}\Big[\frac{\delta(x_1-x_{1,0})}{\sqrt{(x_2-x_{2,0})^2}}\Big]_\oplus. 
\end{eqnarray}

According to Eq.~(\ref{eq:Nor_quasiLCDA}-\ref{eq:local_quasiLCDA}), after adding up all the results of one-loop quasi-DA diagrams, we find the normalized quasi-DA as 
\begin{eqnarray}
\tilde{\phi}(x_1,x_2,\mu)&=&\delta(x_1-x_{1,0})\delta(x_2-x_{2,0})+\frac{\alpha_s  C_F}{8 \pi }\nonumber\\&&\times\Bigg\{\Big[g_2\delta (x_2-x_{2,0})+g_3\delta (x_3-x_{3,0})\nonumber\\&&+\{x_2\leftrightarrow x_1,x_{2,0}\leftrightarrow x_{1,0}\}\Big]\Bigg\}_\oplus,
\end{eqnarray}
with
\begin{widetext}
\begin{align}
g_2=
\left\{ \begin{aligned} 
 &
\frac{\left(x_{1,0}+x_1\right) \left(x_{3,0}+x_3\right) \ln \left(-\frac{x_{3,0}-x_3}{x_3}\right)}{x_{1,0} x_{3,0} \left(x_{3,0}-x_3\right)}-\frac{x_1 \left(2 x_{1,0}+x_{3,0}+x_3\right) \ln \left(-\frac{x_1}{x_3}\right)}{x_{1,0} \left(x_{3,0}-x_3\right) \left(x_{1,0}+x_{3,0}\right)},~x_1<0
\nonumber\\ & 
\frac{x_1 \left(-x_{1,0}+2 x_{2,0}+x_1-2\right)}{\left(x_1-x_{1,0}\right) x_{1,0} \left(x_{2,0}-1\right) \epsilon _{\text{IR}}}+\frac{2 x_1 \ln \left(\frac{4 x_1 \left(x_3-x_{3,0}\right) \left(p^z\right)^2}{\mu ^2}\right)}{x_{1,0} \left(x_3-x_{3,0}\right)}+\frac{x_1 \ln \left(\frac{4 x_1 x_3 \left(p^z\right)^2}{\mu ^2}\right)}{x_{1,0} \left(x_{1,0}+x_{3,0}\right)} \nonumber\\ & +\frac{x_1 \left(-3 x_{1,0}-2 x_{3,0}+x_1\right)}{x_{1,0} \left(x_3-x_{3,0}\right) \left(x_{1,0}+x_{3,0}\right)}-\frac{\left(\left(x_3-x_{3,0}\right){}^2-2 x_3 x_{1,0}\right) \ln \left(\frac{x_3-x_{3,0}}{x_3}\right)}{x_{1,0} \left(x_3-x_{3,0}\right) x_{3,0}},~0<x_1<x_{1,0}
 \nonumber\\ & 
\frac{x_3 \left(-x_{1,0}-2 x_{2,0}+x_1+2\right)}{\left(x_1-x_{1,0}\right) \left(x_{2,0}-1\right) x_{3,0} \epsilon _{\text{IR}}}+\frac{2 x_3 \ln \left(\frac{4 x_3 \left(x_1-x_{1,0}\right) \left(p^z\right)^2}{\mu ^2}\right)}{\left(x_1-x_{1,0}\right) x_{3,0}}+\frac{x_3 \ln \left(\frac{4 x_1 x_3 \left(p^z\right)^2}{\mu ^2}\right)}{x_{3,0} \left(x_{1,0}+x_{3,0}\right)} \nonumber\\ & +\frac{x_3 \left(-2 x_{1,0}-3 x_{3,0}+x_3\right)}{\left(x_1-x_{1,0}\right) x_{3,0} \left(x_{1,0}+x_{3,0}\right)}-\frac{\left(\left(x_1-x_{1,0}\right){}^2-2 x_1 x_{3,0}\right) \ln \left(\frac{x_1-x_{1,0}}{x_1}\right)}{\left(x_1-x_{1,0}\right) x_{1,0} x_{3,0}},~x_{1,0}<x_1<x_{1,0}+x_{3,0}
 \nonumber\\ & 
\frac{\left(x_{1,0}+x_1\right) \left(x_{3,0}+x_3\right) \ln \left(-\frac{x_{3,0}-x_3}{x_3}\right)}{x_{1,0} x_{3,0} \left(x_{3,0}-x_3\right)}-\frac{x_1 \left(2 x_{1,0}+x_{3,0}+x_3\right) \ln \left(-\frac{x_1}{x_3}\right)}{x_{1,0} \left(x_{3,0}-x_3\right) \left(x_{1,0}+x_{3,0}\right)},~x_1>x_{1,0}+x_{3,0}
 \nonumber\\ 
\end{aligned}\right. 
 \end{align} 
 
 \begin{align}
g_3=
\left\{ \begin{aligned} 
 &
\frac{\left(x_{1,0} x_{2,0}+x_1 x_2\right) \ln \left(\frac{x_2-x_{2,0}}{x_2}\right)}{x_{1,0} \left(x_2-x_{2,0}\right) x_{2,0}}-\frac{x_1 \left(x_{1,0}+x_2\right) \ln \left(-\frac{x_1}{x_2}\right)}{x_{1,0} \left(x_2-x_{2,0}\right) \left(x_{1,0}+x_{2,0}\right)},~x_1<0
\nonumber\\ & 
\frac{2 x_1 \left(x_{1,0}+x_2\right)}{\left(x_1+x_2\right) \left(x_1-x_{1,0}\right) x_{1,0} \epsilon _{\text{IR}}}+\frac{x_1 \ln \left(\frac{4 x_1 \left(x_2-x_{2,0}\right) \left(p^z\right)^2}{\mu ^2}\right)}{x_{1,0} \left(x_2-x_{2,0}\right)}+\frac{x_1 \ln \left(\frac{4 x_1 x_2 \left(p^z\right)^2}{\mu ^2}\right)}{x_{1,0} \left(x_{1,0}+x_{2,0}\right)} \nonumber\\ & +\frac{1}{x_1-x_{1,0}}+\frac{2 x_1+x_2}{x_{1,0} \left(x_{1,0}+x_{2,0}\right)}+\frac{\left(x_{1,0} \left(x_{2,0}+x_1\right)-x_1^2\right) \ln \left(\frac{x_2-x_{2,0}}{x_2}\right)}{x_{1,0} \left(x_2-x_{2,0}\right) x_{2,0}},~0<x_1<x_{1,0}
 \nonumber\\ & 
\frac{2 x_2 \left(x_{2,0}+x_1\right)}{\left(x_1+x_2\right) \left(x_2-x_{2,0}\right) x_{2,0} \epsilon _{\text{IR}}}+\frac{x_2 \ln \left(\frac{4 x_2 \left(x_1-x_{1,0}\right) \left(p^z\right)^2}{\mu ^2}\right)}{\left(x_1-x_{1,0}\right) x_{2,0}}+\frac{x_2 \ln \left(\frac{4 x_1 x_2 \left(p^z\right)^2}{\mu ^2}\right)}{x_{2,0} \left(x_{1,0}+x_{2,0}\right)} \nonumber\\ & +\frac{1}{x_2-x_{2,0}}+\frac{x_1+2 x_2}{x_{2,0} \left(x_{1,0}+x_{2,0}\right)}+\frac{\left(\left(x_{1,0}+x_2\right) x_{2,0}-x_2^2\right) \ln \left(\frac{x_1-x_{1,0}}{x_1}\right)}{\left(x_1-x_{1,0}\right) x_{1,0} x_{2,0}},~x_{1,0}<x_1<x_{1,0}+x_{2,0}
 \nonumber\\ & 
\frac{\left(x_{1,0} x_{2,0}+x_1 x_2\right) \ln \left(\frac{x_1-x_{1,0}}{x_1}\right)}{\left(x_1-x_{1,0}\right) x_{1,0} x_{2,0}}-\frac{x_2 \left(x_{2,0}+x_1\right) \ln \left(-\frac{x_2}{x_1}\right)}{\left(x_1-x_{1,0}\right) x_{2,0} \left(x_{1,0}+x_{2,0}\right)},~x_1>x_{1,0}+x_{2,0}.
 \nonumber\\ 
\end{aligned}\right. 
 \end{align} 
 \end{widetext}

There is no divergence for $x_1(x_2)<0$. The expressions of the quasi-DA $\tilde\phi(x_1,x_2,\mu)$ differ by a minus sign between interval $x_1(x_2)<0$ and interval $x_1(x_2)>1-x_2(x_1)$. The above results have infrared divergence in the other two intervals. We can find that this infrared divergence is consistent with the infrared divergence in LCDA Eq.~\eqref{eq:oneloop_Nor_LCDA}, which validate the factorization assumption at one-loop order. A direct demonstration using expansion by region is shown in Appendix C.        

\section{off-shell results}\label{sec:off_shell}

Compared to the continuum space, the renormalization of lattice operators is a necessary ingredient to obtain physical results from numerical simulations. In the literature, it has been noticed that a regularization invariant momentum subtraction method (RI/MOM) \cite{Martinelli:1994ty} can avoid the use of lattice perturbation theory and allow a non-perturbative determination of the renormalization constants of many composite operators \cite{Constantinou:2017sej,Sturm:2009kb,Alexandrou:2017huk,Stewart:2017tvs,Chen:2017mzz,Lin:2017ani,Liu:2018tox,Liu:2019urm,LatticeParton:2018gjr}. In the following we also provide an analysis of the baryon distribution amplitudes in this scheme.

We calculate the quasi-DA in the space-like  $p^2=-\rho{(p^z)}^2<0$ kinematics region.   Fig.~\ref{fig:DAdiagram}(a) gives
\begin{eqnarray}
\tilde{\mathcal{M}}^{a}_{\rm OF}
	&=& ig^2\frac{-C_F}{2} {p^z}^2 \int \frac{d^4 q}{(2\pi)^4} \frac{1}{q^2+i\epsilon}\frac{1 }{(k_3-q)^2+i\epsilon}  \nonumber \\&& \times
	\frac{\delta(x_1p^z-q^z-{k_1}^z)\delta(x_2p^z-{k_2}^z)}{(q+k_1)^2+i\epsilon}  \nonumber \\&&\times\left(-\frac{1}{2}\right){\rm tr}\left(\slashed {p} \gamma^5 \gamma^{\mu} (\slashed{q}+\slashed{k_1})\gamma^5 \gamma^{z}\right)({\slashed{k_3}}-{\slashed{q}})\gamma_{\mu}u(k_3).\nonumber\\
\end{eqnarray}
A subtle issue for the off-shell matrix elements is that there are multiple projection ways, and here we adopt a strategy called the minimal projection~\cite{LatticeParton:2018gjr}. Namely,  we use  the trace formulae technique, and  all kinds of Lorentz structures above the spinor part can be projected out: 
\begin{eqnarray}
	&&\left(-\frac{1}{2}\right){\rm tr}\left(\slashed {p} \gamma^5 \gamma^{\mu} (\slashed{q}+\slashed{k_1})\gamma^5 \gamma^{z}\right)({\slashed{k_3}}-{\slashed{q}})\gamma_{\mu}u(k_3)
	\nonumber\\
	&=&(a_1 + a_2 \slashed{n}_t\slashed{n}_z+ a_3 \slashed{n}_{\perp}\slashed{n}_z+ a_4 \slashed{n}_{\perp}\slashed{n}_t)\tilde{S},
\end{eqnarray}	
where
\begin{widetext}
\begin{eqnarray}	
a_1&=&x_{1,0} \left({(p^z)}^2\rho \left(x_1-x_{1,0}\right)+2
   p\cdot q\right)+{(p^z)}^2 x_1 \rho x_{3,0}+q^2\nonumber\\
a_2&=&\frac{p\cdot q \left(p^0 x_{3,0}-q^0\right)}{p^z}-p^z \Big\{x_{1,0}
   \big[q^0 (1-\rho)+p^0 \left(\rho+1\right) x_{3,0}\big]+x_1
   \left(q^0-p^0 x_{3,0}\right)+p^0 x_{1,0}^2+q^0
   x_{3,0}-p^0 x_1^2\Big\}\nonumber\\
a_3&=&\frac{p\cdot q \left(q_{\perp}-p_{\perp} x_{3,0}\right)}{p^z}+p^z\Big\{x_{1,0} \left(p_{\perp} x_{1,0}+q_{\perp} (1-\rho)\right)+x_{3,0} \big[q_{\perp}+p_{\perp} \left(1+\rho\right) x_{1,0}\big]+x_1 \left(q_{\perp}-p_{\perp}x_{3,0}\right)-p_{\perp} x_1^2\Big\}
  \nonumber\\
a_4&=&\left(x_{1,0}+x_{3,0}+x_1\right) (p_{\perp} q^0-p^0 q_{\perp}).
\end{eqnarray}	
\end{widetext}
In the on-shell limit, the third and the last term $a_3, a_4$ disappear after integrating out the momentum $q$, and the product $\slashed{n}_t\slashed{n}_z$ goes to a unit matrix. Therefore, the summation $a_1+a_2$ captures all terms that lead to UV divergences in the on-shell limit. This is similar to Fig.~\ref{fig:DAdiagram}(b,f,g). The corresponding results are as follows
\begin{widetext}
\begin{align}
&\tilde{\mathcal{M}}_{\rm OF}^a = 
\frac{\delta(x_2-x_{2,0})\alpha_s C_F}{64 \pi (1-\rho)^{3/2} x_{1,0} x_{3,0} (x_{1,0}+x_{3,0}) p^z}\tilde{S}^{(0)}  
\nonumber\\ &\times
\left\{ \begin{aligned} 
 &
\ln \left(\frac{\left(\left(\sqrt{1-\rho }+1\right) \left(x_{1,0}+x_{3,0}\right)-2 x_1\right){}^2}{\left(\left(\sqrt{1-\rho }-1\right) \left(x_{1,0}+x_{3,0}\right)+2 x_1\right){}^2}\right) \left(\rho  \left(x_{1,0}+x_{3,0}\right){}^2 p^0-4 (\rho -1) x_1 \left(x_{1,0}+x_{3,0}\right) p^z-4 x_1^2 p^0\right)\nonumber\\
&
+\ln \left(\frac{\left(\left(\sqrt{1-\rho }+1\right) x_{3,0}+2 x_{1,0}-2 x_1\right){}^2}{\left(\left(\sqrt{1-\rho }-1\right) x_{3,0}-2 x_{1,0}+2 x_1\right){}^2}\right)  \nonumber\\
& \times \left(4 (\rho -1) \left(x_1-x_{1,0}\right) \left(x_{1,0}+x_{3,0}\right) p^z-p^0 \left(\rho  x_{3,0} \left(2 x_{1,0}+x_{3,0}\right)+4 x_1 x_{1,0}-4 x_1^2\right)\right)\nonumber\\
&+\ln \left(\frac{\left(\left(\sqrt{1-\rho }+1\right) x_{1,0}-2 x_1\right){}^2}{\left(\left(\sqrt{1-\rho }-1\right) x_{1,0}+2 x_1\right){}^2}\right) 
\nonumber\\ & \times
\left(4 x_1 \left(x_{3,0} p^0+(\rho -1) \left(x_{1,0}+x_{3,0}\right) p^z\right)-\rho  x_{1,0} \left(x_{1,0}+2 x_{3,0}\right) p^0+4 x_1^2 p^0\right),~x_1<0
\nonumber\\ & 
\ln \left(\frac{\left(\left(\sqrt{1-\rho }+1\right) x_{3,0}+2 x_{1,0}-2 x_1\right){}^2}{\left(\left(\sqrt{1-\rho }-1\right) x_{3,0}-2 x_{1,0}+2 x_1\right){}^2}\right) \nonumber\\
& \times \left(4 (\rho -1) \left(x_1-x_{1,0}\right) \left(x_{1,0}+x_{3,0}\right) p^z-p^0 \left(\rho  x_{3,0} \left(2 x_{1,0}+x_{3,0}\right)+4 x_1 x_{1,0}-4 x_1^2\right)\right)
\nonumber\\ & 
+x_{3,0} p^0 \ln \left(\frac{\left(\sqrt{1-\rho }+1\right)^2}{\left(\sqrt{1-\rho }-1\right)^2}\right) \left(\rho  x_{3,0}+4 x_1\right)-\frac{8 (1-\rho )^{3/2} x_1 x_{3,0} p^0}{z^2},~0<x_1<x_{1,0}
 \nonumber\\ & 
x_{1,0} \ln \left(\frac{\left(\sqrt{1-\rho }+1\right)^2}{\left(\sqrt{1-\rho }-1\right)^2}\right) \left(p^0 \left(\rho  x_{1,0}-4 x_1\right)-4 (\rho -1) \left(x_{1,0}+x_{3,0}\right) p^z\right)
\nonumber\\ & 
+\ln \left(\frac{\left(\left(\sqrt{1-\rho }+1\right) x_{1,0}-2 x_1\right){}^2}{\left(\left(\sqrt{1-\rho }-1\right) x_{1,0}+2 x_1\right){}^2}\right) \left(4 x_1 \left(-x_{3,0} p^0-(\rho -1) \left(x_{1,0}+x_{3,0}\right) p^z\right)+\rho  x_{1,0} \left(x_{1,0}+2 x_{3,0}\right) p^0-4 x_1^2 p^0\right)
\nonumber\\ & 
-\frac{8 (1-\rho )^{3/2} x_{1,0} \left(x_{1,0}+x_{3,0}-x_1\right) p^0}{z^2},~x_{1,0}<x_1<x_{1,0}+x_{3,0}
 \nonumber\\ & 
\ln \left(\frac{\left(\left(\sqrt{1-\rho }+1\right) \left(x_{1,0}+x_{3,0}\right)-2 x_1\right){}^2}{\left(\left(\sqrt{1-\rho }-1\right) \left(x_{1,0}+x_{3,0}\right)+2 x_1\right){}^2}\right) \left(-\rho  \left(x_{1,0}+x_{3,0}\right){}^2 p^0+4 (\rho -1) x_1 \left(x_{1,0}+x_{3,0}\right) p^z+4 x_1^2 p^0\right)
\nonumber\\ & 
+\ln \left(\frac{\left(\left(\sqrt{1-\rho }+1\right) x_{3,0}+2 x_{1,0}-2 x_1\right){}^2}{\left(\left(\sqrt{1-\rho }-1\right) x_{3,0}-2 x_{1,0}+2 x_1\right){}^2}\right)
\nonumber\\ & \times
 \left(p^0 \left(\rho  x_{3,0} \left(2 x_{1,0}+x_{3,0}\right)+4 x_1 x_{1,0}-4 x_1^2\right)-4 (\rho -1) \left(x_1-x_{1,0}\right) \left(x_{1,0}+x_{3,0}\right) p^z\right)
\nonumber\\ & 
+\ln \left(\frac{\left(\left(\sqrt{1-\rho }+1\right) x_{1,0}-2 x_1\right){}^2}{\left(\left(\sqrt{1-\rho }-1\right) x_{1,0}+2 x_1\right){}^2}\right) 
\nonumber\\ & \times
\left(4 x_1 \left(-x_{3,0} p^0-(\rho -1) \left(x_{1,0}+x_{3,0}\right) p^z\right)+\rho  x_{1,0} \left(x_{1,0}+2 x_{3,0}\right) p^0-4 x_1^2 p^0\right),~x_1>x_{1,0}+x_{3,0}
\end{aligned}\right. 
\end{align} 
 
 \begin{align}
&\tilde{\mathcal{M}}_{\rm {OF}}^c = 
\frac{\alpha_s C_F}{16 \pi \sqrt{1-\rho} x_{1,0}} \tilde{S}^{(0)} 
\nonumber\\ & \times 
\left\{ \begin{aligned} 
 &
\Bigg\{\frac{\delta(x_2-x_{2,0})}{ x_1-x_{1,0}}  \Big[ \left(\rho  x_{1,0}-4 x_1\right) \ln \left(\frac{\left(\left(\sqrt{1-\rho }+1\right) x_{1,0}-2 x_1\right){}^2}{\left(\left(\sqrt{1-\rho }-1\right) x_{1,0}+2 x_1\right){}^2}\right)-4 \sqrt{1-\rho } x_{1,0}\Big]\Bigg\}_{\oplus},~x_1<0
\nonumber\\ & 
\Bigg\{\frac{\delta(x_2-x_{2,0})}{x_1-x_{1,0}}  \Big[4 \sqrt{1-\rho } \left(2 x_1-x_{1,0}\right)-\ln \left(\frac{\left(\sqrt{1-\rho }+1\right)^2}{\left(\sqrt{1-\rho }-1\right)^2}\right) \left(4 x_1-\rho  x_{1,0}\right)\Big]\Bigg\}_{\oplus},~0<x_1<x_{1,0}
 \nonumber\\ & 
\Bigg\{\frac{\delta(x_2-x_{2,0})}{x_1-x_{1,0}}  \Big[4 \sqrt{1-\rho } x_{1,0}+\left(4 x_1-\rho  x_{1,0}\right) \ln \left(\frac{\left(\left(\sqrt{1-\rho }+1\right) x_{1,0}-2 x_1\right){}^2}{\left(\left(\sqrt{1-\rho }-1\right) x_{1,0}+2 x_1\right){}^2}\right)\Big]\Bigg\}_{\oplus},~x_1>x_{1,0}
 \nonumber\\ 
\end{aligned}\right. 
 \end{align} 
 
 \begin{align}
&\tilde{\mathcal{M}}_{\rm {OF}}^d = 
\frac{\delta(x_3-x_{3,0})\alpha_s C_F}{8 \pi \sqrt{1-\rho} x_{1,0} x_{2,0} (x_{1,0}+x_{2,0})} \tilde{S}^{(0)}  
\nonumber\\ & \times
\left\{ \begin{aligned} 
 &
x_1 x_{2,0} \ln \left(\frac{\left(\left(\sqrt{1-\rho }+1\right) x_{1,0}-2 x_1\right){}^2}{\left(\left(\sqrt{1-\rho }-1\right) x_{1,0}+2 x_1\right){}^2}\right)\nonumber\\ &
+x_{1,0} \left(x_{1,0}+x_{2,0}-x_1\right) \ln \left(\frac{\left(\left(\sqrt{1-\rho }+1\right) x_{2,0}+2 x_{1,0}-2 x_1\right){}^2}{\left(\left(\sqrt{1-\rho }-1\right) x_{2,0}-2 x_{1,0}+2 x_1\right){}^2}\right),~x_1<0
\nonumber\\ & 
x_1 x_{2,0} \ln \left(\frac{\left(\sqrt{1-\rho }+1\right)^2}{\left(\sqrt{1-\rho }-1\right)^2}\right)
\nonumber\\ &
+x_{1,0} \left(x_{1,0}+x_{2,0}-x_1\right) \ln \left(\frac{\left(\left(\sqrt{1-\rho }+1\right) x_{2,0}+2 x_{1,0}-2 x_1\right){}^2}{\left(\left(\sqrt{1-\rho }-1\right) x_{2,0}-2 x_{1,0}+2 x_1\right){}^2}\right),~0<x_1<x_{1,0}
 \nonumber\\ & 
x_{1,0} \left(x_{1,0}+x_{2,0}-x_1\right) \ln \left(\frac{\left(\sqrt{1-\rho }+1\right)^2}{\left(\sqrt{1-\rho }-1\right)^2}\right)
\nonumber\\ &
-x_1 x_{2,0} \ln \left(\frac{\left(\left(\sqrt{1-\rho }+1\right) x_{1,0}-2 x_1\right){}^2}{\left(\left(\sqrt{1-\rho }-1\right) x_{1,0}+2 x_1\right){}^2}\right),~x_{1,0}<x_1<x_{1,0}+x_{2,0}
 \nonumber\\ & 
x_1 x_{2,0} \left(-\ln \left(\frac{\left(\left(\sqrt{1-\rho }+1\right) x_{1,0}-2 x_1\right){}^2}{\left(\left(\sqrt{1-\rho }-1\right) x_{1,0}+2 x_1\right){}^2}\right)\right)
\nonumber\\ &
-x_{1,0} \left(x_{1,0}+x_{2,0}-x_1\right) \ln \left(\frac{\left(\left(\sqrt{1-\rho }+1\right) x_{2,0}+2 x_{1,0}-2 x_1\right){}^2}{\left(\left(\sqrt{1-\rho }-1\right) x_{2,0}-2 x_{1,0}+2 x_1\right){}^2}\right),~x_1>x_{1,0}+x_{2,0}
 \nonumber\\ 
\end{aligned}\right. 
 \end{align} 
 
 \begin{align}
&\tilde{\mathcal{M}}_{\rm {OF}}^f = 
\frac{\alpha_s C_F}{32 \pi \left(1-\rho\right)^{3/2} x_{3,0} p^z }\tilde{S}^{(0)} 
\nonumber\\ & \times 
\left\{ \begin{aligned} 
& 
\Bigg\{\frac{\delta(x_2-x_{2,0})}{ x_1-x_{1,0}}  \Big[ \ln \left(\frac{\left(\left(\sqrt{1-\rho }+1\right) x_{3,0}+2 x_{1,0}-2 x_1\right){}^2}{\left(\left(\sqrt{1-\rho }-1\right) x_{3,0}-2 x_{1,0}+2 x_1\right){}^2}\right)
\nonumber\\ & \times
 \left(x_{3,0} \left(p^0-(\rho -1) \left(p^0+2 p^z\right)\right)-2 \left(x_1-x_{1,0}\right) \left(p^0-(\rho -1) p^z\right)\right)-\frac{4 (1-\rho )^{3/2} x_{3,0} p^0}{z^2}\Big]\Bigg\}_{\oplus},~x_1<x_{1,0}
 \nonumber\\ & 
\Bigg\{\frac{\delta(x_2-x_{2,0})}{ x_1-x_{1,0}}  \Big[ \frac{4 (1-\rho )^{3/2} \left(-2 x_{1,0}-x_{3,0}+2 x_1\right) p^0}{z^2}-\ln \left(\frac{\left(\sqrt{1-\rho }+1\right)^2}{\left(\sqrt{1-\rho }-1\right)^2}\right)  \nonumber\\ & \times
\left(2 \left(x_1-x_{1,0}\right) \left(p^0-(\rho -1) p^z\right)+x_{3,0} \left((\rho -2) p^0+2 (\rho -1) p^z\right)\right)\Big]\Bigg\}_{\oplus},~x_{1,0}<x_1<x_{1,0}+x_{3,0}
 \nonumber\\ & 
\Bigg\{\frac{\delta(x_2-x_{2,0})}{ x_1-x_{1,0}}  \Big[ \ln \left(\frac{\left(\left(\sqrt{1-\rho }+1\right) x_{3,0}+2 x_{1,0}-2 x_1\right){}^2}{\left(\left(\sqrt{1-\rho }-1\right) x_{3,0}-2 x_{1,0}+2 x_1\right){}^2}\right)
\nonumber\\ & \times \left(2 \left(x_1-x_{1,0}\right) \left(p^0-(\rho -1) p^z\right)+x_{3,0} \left((\rho -2) p^0+2 (\rho -1) p^z\right)\right)+\frac{4 (1-\rho )^{3/2} x_{3,0} p^0}{z^2}\Big]\Bigg\}_{\oplus},~x_1>x_{1,0}+x_{3,0}
 \nonumber\\ 
\end{aligned}\right. 
 \end{align} 
 
 \begin{align}
&\tilde{\mathcal{M}}_{\rm {OF}}^h = 
\frac{\alpha_s C_F}{32 \pi \sqrt{1-\rho} x_{1,0} }\tilde{S}^{(0)} 
\nonumber\\ & \times 
\left\{ \begin{aligned} 
 &
\Bigg\{\frac{\delta(x_2-x_{2,0})-\delta(x_3-x_{3,0})}{ x_{1,0}-x_1}  \Big[ \left(\rho  x_{1,0}-4 x_1\right) \ln \left(\frac{\left(\left(\sqrt{1-\rho }+1\right) x_{1,0}-2 x_1\right){}^2}{\left(\left(\sqrt{1-\rho }-1\right) x_{1,0}+2 x_1\right){}^2}\right)-4 \sqrt{1-\rho } x_{1,0}\Big]\Bigg\}_{\oplus},~x_1<0
\nonumber\\ & 
\Bigg\{\frac{\delta(x_2-x_{2,0})-\delta(x_3-x_{3,0})}{ x_{1,0}-x_1}  4 \sqrt{1-\rho } \left(2 x_1-x_{1,0}\right)+\ln \left(\frac{\left(\sqrt{1-\rho }+1\right)^2}{\left(\sqrt{1-\rho }-1\right)^2}\right) \left(\rho  x_{1,0}-4 x_1\right)\Big]\Bigg\}_{\oplus},~0<x_1<x_{1,0}
 \nonumber\\ & 
\Bigg\{\frac{\delta(x_2-x_{2,0})-\delta(x_3-x_{3,0})}{ x_{1,0}-x_1}  4 \sqrt{1-\rho } x_{1,0}+\left(4 x_1-\rho  x_{1,0}\right) \ln \left(\frac{\left(\left(\sqrt{1-\rho }+1\right) x_{1,0}-2 x_1\right){}^2}{\left(\left(\sqrt{1-\rho }-1\right) x_{1,0}+2 x_1\right){}^2}\right)\Big]\Bigg\}_{\oplus},~x_1>x_{1,0}
 \nonumber\\ 
\end{aligned}\right. 
 \end{align} 
\end{widetext}

\begin{eqnarray}
\tilde{\mathcal{M}}_{\rm {OF}}^{b} &=&\tilde{\mathcal{M}}_{\rm {OF}}^{a} |_{x_2 \leftrightarrow x_1, x_{2,0} \leftrightarrow x_{1,0}},\nonumber\\
\tilde{\mathcal{M}}_{\rm {OF}}^{e} &=&\tilde{\mathcal{M}}_{\rm {OF}}^{c} |_{x_2 \leftrightarrow x_1, x_{2,0} \leftrightarrow x_{1,0}},\nonumber\\
\tilde{\mathcal{M}}_{\rm {OF}}^{g} &=&\tilde{\mathcal{M}}_{\rm {OF}}^{f} |_{x_2 \leftrightarrow x_1, x_{2,0} \leftrightarrow x_{1,0}},\nonumber\\
\tilde{\mathcal{M}}_{\rm {OF}}^{i} &=&\tilde{\mathcal{M}}_{\rm {OF}}^{h} |_{x_2 \leftrightarrow x_1, x_{2,0} \leftrightarrow x_{1,0}},
\end{eqnarray}
 \begin{align}
\tilde{\mathcal{M}}_{\rm {OF}}^{j,k,l} =\tilde{\mathcal{M}}^{j,k,l}. 
 \end{align} 
The subscript “$\rm OF$" indicates the off-shell case. We have found that the self-energy correction of the Wilson line is independent of whether the momentum of the external leg is on-shell or not. Finally, the off-shell quasi-DA   up to one-loop accuracy is given as
\begin{eqnarray}
\tilde{\phi}(x_1,x_2,\mu)_{\rm OF}&=&\delta(x_1-x_{1,0})\delta(x_2-x_{2,0})+\frac{\alpha_s  C_F}{8 \pi }\nonumber\\&&\times\Bigg\{\Big[g^{\prime}_2\delta (x_2-x_{2,0})+g^{\prime}_3\delta (x_3-x_{3,0})\nonumber\\&&+\{x_2\leftrightarrow x_1,x_{2,0}\leftrightarrow x_{1,0}\}\Big]\Bigg\}_\oplus,
\end{eqnarray}
with
\begin{widetext}
\begin{align}
&g^{\prime}_2 = \frac{1}{8 p^z (1-\rho)^{5/2} (x_1-x_{1,0}) x_{1,0} x_{3,0} (x_{1,0}+x_{3,0})}
\nonumber\\ & \times 
\left\{ \begin{aligned} 
 &
-8 (1-\rho )^{3/2} x_{1,0} x_{3,0} \left(x_{1,0}+x_{3,0}\right) \left(p^0+(\rho-1) p^z\right)-(1-\rho) \left(x_1-x_{1,0}\right) \ln \left(\frac{\left(\left(\sqrt{1-\rho }+1\right) \left(x_{1,0}+x_{3,0}\right)-2 x_1\right){}^2}{\left(\left(\sqrt{1-\rho }-1\right) \left(x_{1,0}+x_{3,0}\right)+2 x_1\right){}^2}\right)
\nonumber\\& \times
\Big[-\rho  \left(x_{1,0}+x_{3,0}\right){}^2 p^0+4 (\rho -1) x_1 \left(x_{1,0}+x_{3,0}\right) p^z+4 x_1^2 p^0\Big]-(1-\rho) \ln \left(\frac{\left(\left(\sqrt{1-\rho }+1\right) x_{1,0}-2 x_1\right){}^2}{\left(\left(\sqrt{1-\rho }-1\right) x_{1,0}+2 x_1\right){}^2}\right) 
\nonumber\\& \times
\Big\{2 (\rho -1) x_{3,0} \left(x_{1,0}+x_{3,0}\right) \left(\rho  x_{1,0}-4 x_1\right) p^z-\left(x_1-x_{1,0}\right) \Big[4 x_1 \left(x_{3,0} p^0+(\rho -1) \left(x_{1,0}+x_{3,0}\right) p^z\right)
\nonumber\\&
-\rho  x_{1,0} \left(x_{1,0}+2 x_{3,0}\right) p^0+4 x_1^2 p^0\Big]\Big\}-(1-\rho) \ln \left(\frac{\left(\left(\sqrt{1-\rho }+1\right) x_{3,0}+2 x_{1,0}-2 x_1\right){}^2}{\left(\left(\sqrt{1-\rho }-1\right) x_{3,0}-2 x_{1,0}+2 x_1\right){}^2}\right) 
\nonumber\\& \times
\Big\{2 x_{1,0} \left(x_{1,0}+x_{3,0}\right) \Big[x_{3,0} \left((\rho -2) p^0+2 (\rho -1) p^z\right)+2 \left(x_1-x_{1,0}\right) \left(p^0-\rho  p^z+p^z\right)\Big]
-\left(x_1-x_{1,0}\right) 
 \nonumber\\& \times
\Big[4 \left(x_1-x_{1,0}\right) \left((\rho -1) x_{1,0} p^z+x_1 p^0\right)+2 x_{3,0} \left(2 (\rho -1) x_1 p^z-x_{1,0} \left(\rho  p^0+2 (\rho -1) p^z\right)\right)-\rho  x_{3,0}^2 p^0\Big]\Big\},~x_1<0
\nonumber\\ & 
-8 (1-\rho )^{3/2} x_{3,0} \left(x_{1,0} \left(x_{1,0}+x_{3,0}\right) \left(p^0-(1-\rho) p^z\right)-x_1 \left(x_{1,0} \left(p^0+2 (1-\rho) p^z\right)+2 (1-\rho) x_{3,0} p^z\right)+x_1^2 p^0\right)
\nonumber\\&
+(1-\rho) x_{3,0} \ln \left(\frac{\left(\sqrt{1-\rho }+1\right)^2}{\left(\sqrt{1-\rho }-1\right)^2}\right) \Bigg\{-\rho  x_{1,0} \Bigg[x_{3,0} p^0+2 (\rho -1) \left(x_{1,0}+x_{3,0}\right) p^z\Bigg]+x_1 \Bigg[x_{3,0} \left(\rho  p^0+8 (\rho -1) p^z\right)
\nonumber\\&
-4 x_{1,0} \left(p^0-2 (\rho -1) p^z\right)\Bigg]+4 x_1^2 p^0\Bigg\}+(1-\rho) \ln \left(\frac{\left(\left(\sqrt{1-\rho }+1\right) x_{3,0}+2 x_{1,0}-2 x_1\right){}^2}{\left(\left(\sqrt{1-\rho }-1\right) x_{3,0}-2 x_{1,0}+2 x_1\right){}^2}\right) 
\nonumber\\& \times
\Bigg\{-x_1 \Bigg[2 x_{1,0} x_{3,0} \left((\rho +2) p^0+2 (\rho -1) p^z\right)+\rho  x_{3,0}^2 p^0+4 (\rho -1) x_{1,0}^2 p^z\Bigg]
-4 x_1^2 \Bigg[x_{1,0} \left(2 p^0-\rho  p^z+p^z\right)-(\rho -1) x_{3,0} p^z\Bigg]
\nonumber\\&
+x_{1,0}\Bigg[4 x_{1,0} x_{3,0} \left(2 p^0-\rho  p^z+p^z\right)+x_{3,0}^2 \left(4 \left(p^0+p^z\right)-\rho  \left(p^0+4 p^z\right)\right)+4 x_{1,0}^2 p^0\Bigg]
+4 x_1^3 p^0\Bigg\},~0<x_1<x_{1,0}
 \nonumber\\ & 
-8 (1-\rho )^{3/2} x_{1,0} \left(\left(x_{1,0}+x_{3,0}\right) \left(x_{3,0} \left(p^0+(1-\rho) p^z\right)+x_{1,0} p^0\right)-x_1 x_{3,0} p^0-x_1^2 p^0\right)-(1-\rho) x_{1,0} \ln \left(\frac{\left(\sqrt{1-\rho }+1\right)^2}{\left(\sqrt{1-\rho }-1\right)^2}\right)
\nonumber\\& \times
\Bigg\{2 x_{1,0} x_{3,0} \Bigg[(\rho -4) p^0+2 (\rho -1) p^z\Bigg]+2 x_{3,0}^2 \Bigg[(\rho -2) p^0+2 (\rho -1) p^z\Bigg]+(\rho -4) x_{1,0}^2 p^0+x_1 p^0 \left(4 x_{3,0}-\rho  x_{1,0}\right)+4 x_1^2 p^0\Bigg\}
\nonumber\\&
+(1-\rho) \ln \left(\frac{\left(\left(\sqrt{1-\rho }+1\right) x_{1,0}-2 x_1\right){}^2}{\left(\left(\sqrt{1-\rho }-1\right) x_{1,0}+2 x_1\right){}^2}\right) \Bigg\{2 (\rho -1) x_{3,0} \left(x_{1,0}+x_{3,0}\right) \left(\rho  x_{1,0}-4 x_1\right) p^z
\nonumber\\&
-\left(x_1-x_{1,0}\right) \Bigg[4 x_1 \left(x_{3,0} p^0+(\rho -1) \left(x_{1,0}+x_{3,0}\right) p^z\right)-\rho  x_{1,0} \left(x_{1,0}+2 x_{3,0}\right) p^0+4 x_1^2 p^0\Bigg]\Bigg\},~x_{1,0}<x_1<x_{1,0}+x_{3,0}
 \nonumber\\ & 
8 (1-\rho )^{3/2} x_{1,0} x_{3,0} \left(x_{1,0}+x_{3,0}\right) \left(p^0-(1-\rho) p^z\right)+(1-\rho) \left(x_1-x_{1,0}\right) \ln \left(\frac{\left(\left(\sqrt{1-\rho }+1\right) \left(x_{1,0}+x_{3,0}\right)-2 x_1\right){}^2}{\left(\left(\sqrt{1-\rho }-1\right) \left(x_{1,0}+x_{3,0}\right)+2 x_1\right){}^2}\right)
\nonumber\\& \times
 \Bigg[-\rho  \left(x_{1,0}+x_{3,0}\right){}^2 p^0+4 (\rho -1) x_1 \left(x_{1,0}+x_{3,0}\right) p^z+4 x_1^2 p^0\Bigg]+(1-\rho) \ln \left(\frac{\left(\left(\sqrt{1-\rho }+1\right) x_{1,0}-2 x_1\right){}^2}{\left(\left(\sqrt{1-\rho }-1\right) x_{1,0}+2 x_1\right){}^2}\right)
\nonumber\\& \times
 \Bigg\{2 (\rho -1) x_{3,0} \left(x_{1,0}+x_{3,0}\right) \left(\rho  x_{1,0}-4 x_1\right) p^z-\left(x_1-x_{1,0}\right) \Bigg[4 x_1 \left(x_{3,0} p^0+(\rho -1) \left(x_{1,0}+x_{3,0}\right) p^z\right)
 \nonumber\\&-\rho  x_{1,0} \left(x_{1,0}+2 x_{3,0}\right) p^0+4 x_1^2 p^0\Bigg]\Bigg\} +(1-\rho) \ln \left(\frac{\left(\left(\sqrt{1-\rho}+1\right) x_{3,0}+2 x_{1,0}-2 x_1\right){}^2}{\left(\left(\sqrt{1-\rho }-1\right) x_{3,0}-2 x_{1,0}+2 x_1\right){}^2}\right) \Bigg\{2 x_{1,0} \left(x_{1,0}+x_{3,0}\right)
\nonumber\\&  \times
\Bigg[x_{3,0} \left((\rho -2) p^0+2 (\rho -1) p^z\right)+2 \left(x_1-x_{1,0}\right) \left(p^0-\rho  p^z+p^z\right)\Bigg]-\left(x_1-x_{1,0}\right) \Bigg[4 \left(x_1-x_{1,0}\right) \left((\rho -1) x_{1,0} p^z+x_1 p^0\right)
\nonumber\\& 
-2 x_{3,0} \left(x_{1,0} \left(\rho  p^0+2 (\rho -1) p^z\right)-2 (\rho -1) x_1 p^z\right)-\rho  x_{3,0}^2 p^0\Bigg]\Bigg\},~x_1>x_{1,0}+x_{3,0}
 \nonumber\\ 
\end{aligned}\right. 
\end{align} 
 
 \begin{align}
&g^{\prime}_3 = \frac{1}{4\sqrt{1-\rho}(x_{1,0}+x_{2,0})x_{1,0}x_{2,0}(x_1-x_{1,0})}
\nonumber\\ & \times
\left\{ \begin{aligned} 
 &
4 x_2 \left(x_1-x_{1,0}\right) x_{1,0} \ln \left(\frac{\left(\left(\sqrt{1-\rho }+1\right) x_{2,0}+2 x_{1,0}-2 x_1\right){}^2}{\left(\left(\sqrt{1-\rho }-1\right) x_{2,0}-2 x_{1,0}+2 x_1\right){}^2}\right)+\left(x_{1,0}+x_{2,0}\right) x_{1,0} \left(\rho  x_{2,0}-4 x_2\right) 
\nonumber\\ & \times
\ln \left(\frac{\left(\left(\sqrt{1-\rho }+1\right) x_{2,0}-2 x_2\right){}^2}{\left(\left(\sqrt{1-\rho }-1\right) x_{2,0}+2 x_2\right){}^2}\right)+x_{2,0} \left(\rho  x_{1,0} \left(x_{1,0}+x_{2,0}\right)-4 x_1 \left(2 x_{1,0}+x_{2,0}\right)+4 x_1^2\right) 
\nonumber\\ &\times
\ln \left(\frac{\left(\left(\sqrt{1-\rho }+1\right) x_{1,0}-2 x_1\right){}^2}{\left(\left(\sqrt{1-\rho }-1\right) x_{1,0}+2 x_1\right){}^2}\right),~x_1<0
\nonumber\\ & 
8 \sqrt{1-\rho } x_1 x_{2,0} \left(x_{1,0}+x_{2,0}\right)+4 x_2 \left(x_1-x_{1,0}\right) x_{1,0} \ln \left(\frac{\left(\left(\sqrt{1-\rho }+1\right) x_{2,0}+2 x_{1,0}-2 x_1\right){}^2}{\left(\left(\sqrt{1-\rho }-1\right) x_{2,0}-2 x_{1,0}+2 x_1\right){}^2}\right)
\nonumber\\ &
+\left(x_{1,0}+x_{2,0}\right) x_{1,0} \left(\rho  x_{2,0}-4 x_2\right) \ln \left(\frac{\left(\left(\sqrt{1-\rho }+1\right) x_{2,0}-2 x_2\right){}^2}{\left(\left(\sqrt{1-\rho }-1\right) x_{2,0}+2 x_2\right){}^2}\right)
\nonumber\\ &
+x_{2,0} \ln \left(\frac{\left(\sqrt{1-\rho }+1\right)^2}{\left(\sqrt{1-\rho }-1\right)^2}\right) \left(\rho  x_{1,0} \left(x_{1,0}+x_{2,0}\right)-4 x_1 \left(2 x_{1,0}+x_{2,0}\right)+4 x_1^2\right),~0<x_1<x_{1,0}
 \nonumber\\ & 
-8 \sqrt{1-\rho } x_2 x_{1,0} \left(x_{1,0}+x_{2,0}\right)+x_{1,0} \ln \left(\frac{\left(\sqrt{1-\rho }+1\right)^2}{\left(\sqrt{1-\rho }-1\right)^2}\right) \left(4 x_1 \left(x_2-x_{2,0}\right)-(\rho -4) x_{2,0} \left(x_{1,0}+x_{2,0}\right)\right)
\nonumber\\ &-x_{2,0} \left(\rho  x_{1,0} \left(x_{1,0}+x_{2,0}\right)-4 x_1 \left(2 x_{1,0}+x_{2,0}\right)+4 x_1^2\right) \ln \left(\frac{\left(\left(\sqrt{1-\rho }+1\right) x_{1,0}-2 x_1\right){}^2}{\left(\left(\sqrt{1-\rho }-1\right) x_{1,0}+2 x_1\right){}^2}\right),~x_{1,0}<x_1<x_{1,0}+x_{2,0}
 \nonumber\\ & 
-4 \left(x_1-x_{1,0}\right) x_{1,0} \left(x_{1,0}+x_{2,0}-x_1\right) \ln \left(\frac{\left(\left(\sqrt{1-\rho }+1\right) x_{2,0}+2 x_{1,0}-2 x_1\right){}^2}{\left(\left(\sqrt{1-\rho }-1\right) x_{2,0}-2 x_{1,0}+2 x_1\right){}^2}\right) 
\nonumber\\ &
+\frac{\left(x_1-x_{1,0}\right) x_{1,0} \left(x_{1,0}+x_{2,0}\right) \left(\rho  x_{2,0}-4 x_2\right) \ln \left(\frac{\left(\left(\sqrt{1-\rho }+1\right) x_{2,0}-2 x_2\right){}^2}{\left(\left(\sqrt{1-\rho }-1\right) x_{2,0}+2 x_2\right){}^2}\right)}{x_2-x_{2,0}}
\nonumber\\ &
-x_{2,0} \left(\rho  x_{1,0} \left(x_{1,0}+x_{2,0}\right)-4 x_1 \left(2 x_{1,0}+x_{2,0}\right)+4 x_1^2\right) \ln \left(\frac{\left(\left(\sqrt{1-\rho }+1\right) x_{1,0}-2 x_1\right){}^2}{\left(\left(\sqrt{1-\rho }-1\right) x_{1,0}+2 x_1\right){}^2}\right),~x_1>x_{1,0}+x_{2,0}.
 \nonumber\\ 
\end{aligned}\right. 
 \end{align} 
 \end{widetext}

\section{matching kernel}
\label{sec:matching}

In the large momentum $p^z \gg \Lambda_{\rm QCD}$ limit, the quasi observables  can be factorized as a convolution of a perturbatively calculable matching coefficient and the corresponding light-cone observable  up to power corrections suppressed by $\big(\frac{1}{x_1p^z},\frac{1}{x_2p^z},\frac{1}{(1-x_1-x_2)p^z}\big)$. Through this factorization, one can extract light-cone observables from quasi-ones calculated on the lattice. The matching of quasi-DA and LCDA is  given as
\begin{eqnarray}
\label{DAmatching}
\tilde{\Phi}(x_1,x_2,\mu) &=& \int dy_1dy_2 \mathcal{C}(x_1,x_2,y_1,y_2,\mu)\Phi(y_1,y_2,\mu)\nonumber\\&&
+\mathcal{O}\Big(\frac{1}{x_1p^z},\frac{1}{x_2p^z},\frac{1}{(1-x_1-x_2)p^z}\Big).\nonumber\\
\end{eqnarray}

With the results presented in the previous sections, one can easily obtain  the matching kernel in the ${\overline {\rm MS}}$ scheme up to one-loop level
\begin{eqnarray}
\mathcal{C}(x_1,x_2,y_1,y_2,\mu)&=&\delta(x_1-y_1)\delta(x_2-y_2)+\frac{\alpha_s  C_F}{8 \pi }\nonumber\\&&\times\Bigg[C_2(x_1,x_2,y_1,y_2)\delta (x_2-y_2)\nonumber\\&&+C_3(x_1,x_2,y_1,y_2)\delta (x_3-y_3)\nonumber\\&&+\{x_1\leftrightarrow x_2,y_1\leftrightarrow y_2\}\Bigg]_\oplus,\nonumber\\
\end{eqnarray}

where $y_3=1-y_1-y_2$ and
\begin{widetext}
\begin{eqnarray}
&&C_2(x_1,x_2,y_1,y_2)= \nonumber \\&&
\left\{ \begin{aligned}
&\frac{\left(x_1+y_1\right) \left(x_3+y_3\right) \ln
   \frac{y_1-x_1}{-x_1}}{y_1 \left(y_1-x_1\right) y_3}-\frac{x_3
   \left(x_1+y_1+2 y_3\right) \ln
    \frac{x_3}{-x_1} }{\left(y_1-x_1\right) y_3
   \left(y_1+y_3\right)}       ,~x_1<0\\
&\frac{\left(x_1-3
    y_1-2 y_3\right) x_1}{y_1 \left(x_{3}-y_3\right)
   \left(y_1+y_3\right)}-\frac{\left[\left(x_3-y_3\right)^2-2 x_3 y_1\right]
   \ln \frac{x_3-y_3}{x_3}}{y_1 \left(x_3-y_3\right) y_3}+\frac{2 x_1 \ln \frac{4 x_1  \left(x_3-y_3\right) p_z^2}{\mu^2}}{y_1 \left(x_3-y_3\right)}+\frac{x_1 \ln
    \frac{4 x_1 x_3 p_z^2}{\mu^2} }{y_1 \left(y_1+y_3\right)}       ,~0<x_1<y_1\\
&\frac{\left(x_3-2
   y_1-3 y_3\right) x_3}{y_3 \left(x_{1}-y_1\right)
   \left(y_1+y_3\right)} -\frac{\left[\left(x_1-y_1\right)^2-2 x_1 y_3\right]
   \ln \frac{x_1-y_1}{x_1}}{\left(x_1-y_1\right) y_1 y_3} + \frac{2 x_3 \ln \frac{4 x_3
   \left(x_1-y_1\right) p_z^2}{\mu^2}}{\left(x_1-y_1\right) y_3}+\frac{x_3 \ln
   \frac{4 x_1 x_3 p_z^2}{\mu^2}}{y_3 \left(y_1+y_3\right)}       ,~y_1<x_1<y_1+y_3\\
&\frac{\left(x_1+y_1\right) \left(x_3+y_3\right) \ln
    \frac{y_3-x_3}{-x_3} }{ y_1 y_3 \left(y_3-x_3\right)}-\frac{x_1
   \left(x_3+2 y_1+y_3\right) \ln  \frac{x_1}{-x_3} }{ y_1
   \left(y_3-x_3\right) \left(y_1+y_3\right)},~x_1>y_1+y_3
\nonumber\\
\end{aligned}\right.\nonumber\\   
&&C_3(x_1,x_2,y_1,y_2)= \\&&
\left\{ \begin{aligned}
&\frac{\left(x_1 x_2+y_1 y_2\right) \ln
    \frac{x_2-y_2}{x_2} }{y_1 \left(x_2-y_2\right) y_2}-\frac{x_1
   \left(x_2+y_1\right) \ln \frac{-x_1}{x_2}}{y_1 \left(x_2-y_2\right)
   \left(y_1+y_2\right)},~x_1<0\\
&\frac{1}{x_1-y_1}+\frac{2 x_1+x_2}{y_1
   \left(y_1+y_2\right)}+\frac{\left[ \left(x_1+y_2\right)y_1-x_1{}^2\right] \ln
    \frac{x_2-y_2}{x_2} }{y_1 \left(x_2-y_2\right) y_2}+\frac{x_1 \ln  \frac{4 x_1
   \left(x_2-y_2\right) p_z^2}{\mu^2} }{y_1 \left(x_2-y_2\right)}+\frac{x_1 \ln
    \frac{4 x_1 x_2 p_z^2}{\mu^2} }{y_1
   \left(y_1+y_2\right)},~0<x_1<y_1\\
&\frac{1}{x_2-y_2}+\frac{  x_1+2 x_2 }{y_2
   \left(y_1+y_2\right)}+\frac{ \left[\left(x_2+y_1\right) y_2-x_2{}^2\right] \ln
   \frac{x_1-y_1}{x_1}}{\left(x_1-y_1\right) y_1 y_2}+\frac{ x_2 \ln \frac{4 x_2
   \left(x_1-y_1 \right)p_z^2}{\mu^2}}{\left(x_1-y_1\right) y_2}+\frac{ x_2 \ln
   \frac{4 x_1 x_2 p_z^2}{\mu^2}}{y_2
   \left(y_1+y_2\right)},~y_1<x_1<y_1+y_2\\
&\frac{\left(x_1 x_2+y_1 y_2\right) \ln
   \frac{x_1-y_1}{x_1}}{y_1 \left(x_1-y_1\right) y_2}- \frac{x_2 \left(x_1+y_2\right) \ln
   \frac{-x_2}{x_1}}{y_2 \left(x_1-y_1\right)
   \left(y_1+y_2\right)},~x_1>y_1+y_2.\nonumber\\
\end{aligned}\right.\nonumber\\
\end{eqnarray}
\end{widetext}
If we integrate over the physical region of the momentum fraction $y_{1,2}$ of the matching kernel, we find that the integral diverges. In order to eliminate this ultraviolet divergence and renormalize the lattice operators, we need a suitable renormalization of the quasi-DA $\tilde{\phi}(x_1,x_2,\mu)$. In the RI/MOM scheme, this is given as
\begin{eqnarray}
\label{eq:norquasiRIMOMLCDA}
\tilde{\phi}(x_1,x_2,\mu)_{\rm RI/MOM}=\frac{\tilde{\phi}(x_1,x_2,\mu)}{\tilde{\phi}(x_1,x_2,\mu)_{\rm OF}}.
\end{eqnarray} 
Therefore, the renormalized matching coefficient $\mathcal{C}$ in Eq.~\eqref{DAmatching} is
\begin{eqnarray}
\mathcal{C}^{\mathcal {R}}(x_1,x_2,y_1,y_2,\mu)&=&\delta(x_1-y_1)\delta(x_2-y_2)+\frac{\alpha_s  C_F}{8 \pi }\nonumber\\&&\times\Bigg[C^{\prime}_2(x_1,x_2,y_1,y_2)\delta (x_2-y_2)\nonumber\\&&+C^{\prime}_3(x_1,x_2,y_1,y_2)\delta (x_3-y_3)\nonumber\\&&+\{x_1\leftrightarrow x_2,y_1\leftrightarrow y_2\}\Bigg]_\oplus,
\end{eqnarray}
where $C^{\prime}_2=C_2-g^{\prime}_2|_{x_{1,0} \rightarrow y_1, x_{2,0} \rightarrow y_2}$ and $C^{\prime}_3=C_3-g^{\prime}_3|_{x_{1,0} \rightarrow y_1, x_{2,0} \rightarrow y_2}$. In the RI/MOM scheme, the UV divergence in the quasi-DAs can be removed by the renormalization constant determined nonperturbatively.

It should be emphasized that although in the above calculation, an off-shell result is used to remove the UV divergences in the integration, additional infrared effects are likely to be introduced. Recently, hybrid renormalization and self-renormalization schemes have been adopted to obtain a more coherent result~\cite{Ji:2020brr,LatticePartonCollaborationLPC:2021xdx,Chou:2022drv,Zhang:2022xuw}. The hybrid renormalization scheme treats the short-distance and long-distance renormalization separately while the self-renormalization scheme aims to extract the linear divergence by the zero-momentum matrix element. An analysis of the renormalization of the baryon quasi-DA in such a scheme is undergoing.  More recently, a newly proposed method is also shown in Ref.~\cite{Constantinou:2022aij}. 
 
\section{Summary}\label{sec:summary}
In this work, we have pointed out that  LCDAs of a light baryon can be obtained through a simulation of a quasi-distribution amplitude calculable on lattice QCD under the framework of large-momentum effective theory. We have calculated the one-loop perturbative contributions to LCDA and quasi-distribution amplitudes and explicitly have demonstrated the factorization of quasi-distribution amplitudes at the one-loop level. A direct analysis using expansion by region also verifies the factorizability of quasi-DA. Based on the perturbative results, we have derived the matching kernel.

For the renormalization of quasi-distribution amplitudes,  we have adopted the simplest procedure at this stage and subtracted the results with an off-shell parton state as a RI/MOM result.  Our result provides a first step to obtaining the LCDA from first principle lattice QCD calculations in the future. An improved renormalization procedure might be performed in the self-renormalization or hybrid approach.

\section*{Acknowledgment}

We thank  Minhuan Chu, Jun Hua, Xiangdong Ji,  Yushan Su and Qi-An Zhang for their valuable discussions.     This work is supported in part by the Natural Science Foundation of China under Grants No. 12205180, No. 12147140, No. 11735010, No. 12125503, and No. 11905126, by the Natural Science Foundation of Shanghai, by the Project funded by China Postdoctoral Science Foundation under Grant No. 2022M712088.

\begin{appendix} 

\section{Gauge invariance in LCDAs}\label{identity_wilsonline}
According to Eqs.~(\ref{eq:definitionLCDA}-\ref{eq:wilsonline}), a  gauge-invariant form for the LCDA of a light baryon $\Lambda$ 
can be constructed as:
\begin{eqnarray} 
 &&\epsilon_{ijk}  \langle 0|\mathcal{W}_{ii^{\prime}}(\infty,t_1 n) u_{i^{\prime}}^T(t_1 n)  \Gamma \mathcal{W}_{jj^{\prime}}(\infty,t_2 n) d_{j^{\prime}}(t_2 n) \nonumber\\& &\times 
 \mathcal{W}_{kk^{\prime}}(\infty,0) s_{k^{\prime}}(0) |\Lambda\rangle .
\end{eqnarray}

If we focus on the color structure, we can find 
\begin{eqnarray}\label{eq:GILCDA}
&&\epsilon_{ijk}  \mathcal{W}_{ii^{\prime}}(\infty,t_1 n) \mathcal{W}_{jj^{\prime}}(\infty,t_2 n) \mathcal{W}_{kk^{\prime}}(\infty,0)\nonumber\\
&=& \epsilon_{ijk} \mathcal{W}_{il}(\infty,0)\mathcal{W}_{li^{\prime}}(0,t_1n) \mathcal{W}_{jm}(\infty,0)\mathcal{W}_{mj^{\prime}}(0,t_2n) \nonumber\\&&\times \mathcal{W}_{kk^{\prime}}(\infty,0)\nonumber\\
&=& \mathcal{W}_{li^{\prime}}(0,t_1n) \mathcal{W}_{mj^{\prime}}(0,t_2n) \epsilon_{lmk^{\prime}}{\rm{det}}\left|\mathcal{W}(\infty,0)\right|.
\end{eqnarray}
Here we have used the identity, the definition of $3 \times 3$ matrix determinant, $\epsilon_{ijk}  \mathcal{W}_{il}(\infty,0) \mathcal{W}_{jm}(\infty,0) \mathcal{W}_{kk^{\prime}}(\infty,0)=\epsilon_{lmk^{\prime}}{\rm{det}}\left|\mathcal{W}(\infty,0)\right|$. The Wilson line satisfy the property of $SU(3)$ group. Therefore, the gauge-invariant LCDA Eq.~\eqref{eq:GILCDA} can be also written as
\begin{eqnarray} 
\epsilon_{lmk^{\prime}}  \langle 0|\mathcal{W}_{li^{\prime}}(0,t_1 n) u_{i^{\prime}}^T(t_1 n)  \Gamma \mathcal{W}_{mj^{\prime}}(0,t_2 n) d_{j^{\prime}}(t_2 n) s_{k^{\prime}}(0) |\Lambda\rangle \nonumber,
\end{eqnarray}
or equivalently:
\begin{eqnarray}
\epsilon_{ijk}  \langle 0|\mathcal{W}_{ii^{\prime}}(0,t_1 n) u_{i^{\prime}}^T(t_1 n)  \Gamma \mathcal{W}_{jj^{\prime}}(0,t_2 n) d_{j^{\prime}}(t_2 n) s_{k}(0) |\Lambda\rangle. \nonumber\\
\end{eqnarray}

\section{Projection and Trace formulae}\label{App:trace_formulae}

We consider a tree-level matrix element:
\begin{eqnarray}
&&\int \frac{da_1p^{+}}{2\pi}\int \frac{da_2p^{+}}{2\pi} e^{i(x_{1}a_1+x_{2}a_2)p^+} \frac{\epsilon_{ijk}\epsilon_{abc}}{6}\langle 0|u_{i}^T(a_1 n) \nonumber\\
&&\times C \slashed{n} \gamma^5 d_{j}(a_2 n) s_{k}(0) | u_{a}\left(x_{1,0} p\right) d_{b}(x_{2,0} p) s_{c}(x_{3,0}p)\rangle \nonumber\\
&=&(p^+)^2\langle 0 |
u^T(a_1 n)  C \slashed{n} \gamma^5 d(a_2 n) s(0) \frac{1}{\sqrt{2}}[  b_{\uparrow,u}^{\dagger}(x_{1,0}p )\nonumber\\
&&\times b_{\downarrow,d}^{\dagger}\left(x_{2,0} P\right)- b_{\downarrow,u}^{\dagger}\left(x_{1,0} p\right)b_{\uparrow,d}^{ \dagger}\left(x_{2,0} p\right) ] b_{s}^{\dagger}\left(x_{3,0} p\right)  | 0 \rangle  \nonumber\\
&=&-\delta(x_1-x_{1,0})\delta(x_2-x_{2,0})\frac{1}{\sqrt{2}}{\rm{Tr}} \{ [ u_{\downarrow} \left(x_{2,0} p\right) u^T_{\uparrow} \left(x_{1,0} p\right)\nonumber\\
&&- u_{\uparrow}\left(x_{2,0} p\right) u_{\downarrow}\left(x_{1,0} p\right) ]C \slashed{n} \gamma^5 \} u\left(x_{3,0} p\right) , 
\end{eqnarray}
where the arrow $\uparrow$ and $\downarrow$ denote the spin $+1/2$ and $-1/2$ for $ud$ quark pair. Using the spinor 
\begin{eqnarray}
u^{\uparrow}(x p)=\sqrt{x p^z}\left(
\begin{array}{ccc}
1\\ 0\\1\\ 0
\end{array}
\right), u^{\downarrow}(x p)= \sqrt{x p^z}\left(
\begin{array}{ccc}
0\\1\\ 0 \\-1
\end{array}
\right)
\end{eqnarray} 
under the Dirac representation, one has  
\begin{eqnarray}
 \frac{ u_{\downarrow} \left(x_{2,0} p\right) u^T_{\uparrow} \left(x_{1,0} p\right)- u_{\uparrow}\left(x_{2,0} p\right) u^T_{\downarrow}\left(x_{1,0} p\right)}{\sqrt{2}}  = c_1\frac{1}{2} \slashed{p} C \gamma^5, \nonumber\\
\end{eqnarray}
with the coefficient $c_1= -\sqrt{2{x_{1,0} x_{2,0}}}$. 

Since this factor $c_1$ appears both in the evaluation of tree-level and one-loop operator matrix elements, one can neglect this factor. Thus one can employ a tree-level operator matrix element
\begin{eqnarray}
&&\int \frac{d a_1 p^{+}}{2\pi}\int \frac{d a_2p^{+}}{2\pi} e^{i(x_1a_1+x_2a_2)p^+} \frac{\epsilon_{ijk}\epsilon_{abc}}{6}\langle 0|u_{i}^T(a_1 n) \nonumber\\
&&\times  C \slashed{n} \gamma^5 d_{j}(a_2 n) s_{k}(0) | u_{a}\left(x_{1,0} p\right) d_{b}(x_{2,0} p) s_{c}(x_{3,0}p)\rangle|_{\rm tree}\nonumber\\
&=&\delta(x_1-x_{1,0})\delta(x_2-x_{2,0})\frac{1}{2}{\rm{tr}} \left\{ \gamma^5 \slashed{p}\slashed{n}\gamma^5  \right\} u\left(x_{3,0} p\right) \nonumber\\
&=& 2p^+\delta(x_1-x_{1,0})\delta(x_2-x_{2,0}) u\left(x_{3,0} p\right).
\end{eqnarray}

The normalization of the LCDA will lead to the local operator matrix element
\begin{eqnarray}
&&\int dx_1\int dx_2\int \frac{da_1p^+}{2\pi}\int \frac{da_2p^+}{2\pi} e^{i(x_{1}a_1+x_{2}a_2)p^+}\nonumber\\
&&\times\frac{\epsilon_{ijk}\epsilon_{abc}}{6}\left\langle 0\left|
u_{i}^T(a_1 n)  C \slashed{n} \gamma^5 d_{j}(a_2 n) s_{k}(0) \right| u_{a} d_{b}s_{c}\right\rangle\nonumber\\
&=&(p^+)^2\int \frac{da_1}{2\pi} \int \frac{da_2}{2\pi} \delta(a_1 p^+) \delta(a_2 p^+)  \nonumber\\
&&\times\frac{\epsilon_{ijk}\epsilon_{abc}}{6}\left\langle 0\left|
u_{i}^T(a_1 n)  C \slashed{n} \gamma^5 d_{j}(a_2 n) s_{k}(0) \right| u_{a} d_{b} s_{c}\right\rangle\nonumber\\
&=& \frac{\epsilon_{ijk}\epsilon_{abc}}{6}\left\langle 0\left|
u_{i}^T(0)  C \slashed{n} \gamma^5 d_{j}(0) s_{k}(0) \right| u_{a} d_{b} s_{c} \right\rangle\nonumber\\
&=& S.
\end{eqnarray}

Therefore the normalized LCDA at the one-loop accuracy is
\begin{eqnarray}
\label{app:nor_LCDA}
\phi&=& \frac{S^{(0)}\delta(x_1-x_{10})\delta(x_2-x_{20}) +\sum_{i}\mathcal{M}^{i}}{S^{(0)}+S^{(1)}}\nonumber\\
&=&\delta(x_1-x_{1,0})\delta(x_2-x_{2,0})+\frac{1}{S^{(0)}}\Big(\sum_{i}\mathcal{M}^{i}\nonumber\\&&-\delta(x_1-x_{1,0})\delta(x_2-x_{2,0})S^{(1)}\Big)
\nonumber\\
&=&\delta(x_1-x_{1,0})\delta(x_2-x_{2,0})+\frac{1}{S^{(0)}}\Big(\sum_{i}\mathcal{M}^{i}\nonumber\\&&-\delta(x_1-x_{1,0})\delta(x_2-x_{2,0})\nonumber\\
&& \times \int dy_1\int dy_2\sum_{i}\mathcal{M}^{i}|_{x_1\to y_1,x_2\to y_2}\Big),
\end{eqnarray}
where we adopted the convention for perturbative expansion. The normalization of quasi-DA will give a similar form.

\section{Expansion by regions and Factorization of quasi-DA at one-loop}\label{App:Expansion_by_region}

In LaMET, 
it is conjectured that the quasi-distribution amplitudes can be factorized as a convolution of the LCDAs and a hard kernel. A rigorous proof of quasi PDFs can be elegantly found in Refs.~\cite{Ma:2014jla,Izubuchi:2018srq}. In the following, we adopt the technique of expansion by region~\cite{Beneke:1997zp} for quasi-DA and explicitly demonstrate the factorization of quasi-DA. 

In the definition of quasi-DA, one has two popular choices for the Lorentz structures in the
interpolating operator: $\tilde\Gamma=C\gamma^5\gamma^z$, and $\tilde\Gamma=C\gamma^5\gamma^t$. We will show that the short-distance results,  namely the hard kernel,
are the same at the one-loop level in $\overline {\rm MS}$ scheme.

We will analyze the normalized coefficient $\tilde{\phi}^{i}|_{(1/0)}$ or $\mathcal{M}^{i}|_{(1/0)}$:
\begin{eqnarray}
\tilde{\phi}^{i}&=&\tilde{\phi}^{i}|_{(1/0)}\tilde{\mathcal{S}},
\end{eqnarray}
\begin{eqnarray}
\mathcal{M}^{i}&=&\mathcal{M}^{i}|_{(1/0)}\mathcal{S}.
\end{eqnarray}

In the quasi-DA, there are three potential leading power contributions according to the decomposition of the momentum $q=(q^+,q^-,q_\perp)$, \begin{itemize}
\item Hard mode  with  $q\sim(1,1,1)Q$:

In this region, all the hard kinetic components must be retained. Then, one can find that the magnitude of the amplitude is order one $\mathcal O(1)$.

\item Collinear mode with  $q\sim(Q,\Lambda^2_{\rm QCD}/Q,\Lambda_{\rm QCD})$: 

In this region, one can find that the amplitude is $\mathcal O(1)$, and actually the amplitudes for both structures are reduced to the LCDA: 
\begin{widetext}
\begin{eqnarray}
\tilde{\phi}^{a}_{(1/0)}|_{C}&=&ig^2\frac{C_F}{2} p^z\delta(x_2-x_{2,0})\int \frac{d^4 q}{(2\pi)^4} \frac{\delta(x_1p^z-q^z-{k_1}^z)}{(q+k_1)^2+i\epsilon} \frac{1}{q^2+i\epsilon}\frac{q_\perp^2 }{(k_3-q)^2+i\epsilon}|_{C}
\nonumber\\
&=&ig^2\frac{C_F}{2} \frac{p^+-p^-}{\sqrt{2}}\delta(x_2-x_{2,0})\int \frac{d^4 q}{(2\pi)^4} \frac{\sqrt{2}\delta\big((x_1p^+-q^+-{k_1}^+)-(x_1p^--q^--{k_1}^-)\big)}{(q+k_1)^2+i\epsilon}\nonumber\\
&&\frac{1}{q^2+i\epsilon}\frac{q_\perp^2 }{(k_3-q)^2+i\epsilon}|_{C}\nonumber\\
&=&ig^2\frac{C_F}{2} p^+\delta(x_2-x_{2,0})\int \frac{d^4 q}{(2\pi)^4} \frac{\delta\big(x_1p^+-q^+-{k_1}^+\big)}{(q+k_1)^2+i\epsilon}\frac{1}{q^2+i\epsilon}\frac{q_\perp^2 }{(k_3-q)^2+i\epsilon}\nonumber\\
&=&\phi^{a}_{(1/0)}.\nonumber\\
\end{eqnarray}
\begin{eqnarray}
\tilde{\phi}^{c}_{(1/0)}|_{C}
&=&ig^2C_F p^z\left[\delta(x_2-x_{2,0})\int \frac{d^4 q}{(2\pi)^4} \frac{\delta(x_1p^z-q^z-{k_1}^z)}{(k_1+q)^2+i\epsilon} \frac{1}{q^2+i\epsilon}\frac{2k_1^z+q^0+q^z}{q^z}\right]_\oplus|_{C}
\nonumber\\
&=&ig^2C_F \frac{p^+}{\sqrt{2}}\left[\delta(x_2-x_{2,0})\int \frac{d^4 q}{(2\pi)^4} \frac{\delta(\frac{x_1p^+}{\sqrt{2}}-\frac{q^+-q^-}{\sqrt{2}}- \frac{k_1^+}{\sqrt{2}})}{(k_1+q)^2+i\epsilon} \frac{1}{q^2+i\epsilon}\frac{2\frac{k_1^+}{\sqrt{2}}+\frac{q^++q^-}{\sqrt{2}}+\frac{q^+-q^-}{\sqrt{2}}}{\frac{q^+-q^-}{\sqrt{2}}}\right]_\oplus|_{C}
\nonumber\\
&=&ig^2C_F \frac{p^+}{\sqrt{2}}\left[\delta(x_2-x_{2,0})\int \frac{d^4 q}{(2\pi)^4} \frac{\sqrt{2}\delta(x_1p^+-q^+- k_1^+)}{(k_1+q)^2+i\epsilon} \frac{1}{q^2+i\epsilon}\frac{2k_1^++q^++q^+}{q^+}\right]_\oplus
\nonumber\\
&=&ig^2C_F p^+\left[\delta(x_2-x_{2,0})\int \frac{d^4 q}{(2\pi)^4} \frac{\delta[(x_1-x_{1,0})p^+ -q^+]}{(k_1+q)^2+i\epsilon} \frac{1}{q^2+i\epsilon}\frac{2(k_1^++q^+)}{q^+}\right]_\oplus
\nonumber\\
&=&\phi^c_{(1/0)}
\nonumber.
\end{eqnarray}
\end{widetext}

\item Soft mode  $q\sim(\Lambda_{\rm QCD},\Lambda_{\rm QCD},\Lambda_{\rm QCD})$: 

In this kinematics region, one can find the power of the amplitude is $\mathcal O(\Lambda_{\rm QCD}/Q)$, and namely, this amplitude is suppressed. 
\end{itemize}

This analysis indicates that the amplitude from  Fig.~\ref{fig:DAdiagram}(a) and \ref{fig:DAdiagram}(c) are independent of the Lorentz structure, and moreover, we have checked other amplitudes in Fig.~\ref{fig:DAdiagram}. We have found that the one-loop LCDA and quasi-DA for baryon $\Lambda$ does not contain the soft contributions. The one-loop quasi-DA contain the collinear and hard mode. As anticipated the one-loop LCDA only contains the collinear mode at leading power. As a result,  QCD factorization shows that the hard and collinear modes in the quasi-DA can be factorized  into a convolution of the hard matching coefficient and the LCDA which only contains collinear modes.

\end{appendix}

\end{document}